\documentclass[twocolumn]{aastex701}

\usepackage{xcolor}
\newcommand{\ca}{Ca~{\sc ii}}
\newcommand{\sunriseiii}{{\sc Sunrise~iii}}

\begin{document}
\title{\sunriseiii: Instrument, mission, data, and first results}

%\input{authors}
%!TEX root = main.tex
\author[orcid=0000-0002-3418-8449,sname='Solanki']{Sami~K.~Solanki} \affiliation{Max-Planck-Institut für Sonnensystemforschung, Justus-von-Liebig-Weg 3, 37077 Göttingen, Germany}\email{solanki@mps.mpg.de}		
\author[orcid=0000-0003-3490-6532,sname='Smitha']{H.~N.~Smitha} \affiliation{Max-Planck-Institut für Sonnensystemforschung, Justus-von-Liebig-Weg 3, 37077 Göttingen, Germany}\email{narayanamurthy@mps.mpg.de}		

%%% Lead-CoIs		mandatory for all papers
\author[orcid=0000-0003-1459-7074,sname='Lagg']{Andreas~Lagg} \affiliation{Max-Planck-Institut für Sonnensystemforschung, Justus-von-Liebig-Weg 3, 37077 Göttingen, Germany}\email{lagg@mps.mpg.de}		
\author[orcid=0000-0002-9972-9840,sname='Gandorfer']{Achim~Gandorfer} \affiliation{Max-Planck-Institut für Sonnensystemforschung, Justus-von-Liebig-Weg 3, 37077 Göttingen, Germany}\email{gandorfer@mps.mpg.de}		
\author[orcid=0000-0002-3387-026X,sname='del~Toro~Iniesta']{Jose~Carlos~del~Toro~Iniesta} \affiliation{Instituto de Astrofísica de Andalucía, CSIC, Glorieta de la Astronomía s/n, 18008 Granada, Spain}\affiliation{Spanish Space Solar Physics Consortium}\email{jti@iaa.es}		
\author[orcid=0000-0002-5054-8782,sname='Katsukawa']{Yukio~Katsukawa} \affiliation{National Astronomical Observatory of Japan, 2-21-1 Osawa, Mitaka, Tokyo 181-8588, Japan}\affiliation{Department of Astronomy, The University of Tokyo, 7-3-1, Hongo, Bunkyo-ku, Tokyo 113-0033, Japan}\affiliation{Department of Astronomical Science, The Graduate University for Advanced Studies (SOKENDAI), 2-21-1 Osawa, Mitaka, Tokyo 1818588, Japan}\email{yukio.katsukawa@nao.ac.jp} 
%\author[orcid=0000-0002-5054-8782,sname='Katsukawa']{Yukio~Katsukawa} \affiliation{National Astronomical Observatory of Japan, 2-21-1 Osawa, Mitaka, Tokyo 181-8588, Japan}\affiliation{Department of Earth and Planetary Science, The University of Tokyo, 7-3-1, Hongo, Bunkyo-ku, Tokyo 113-0033, Japan}\affiliation{Department of Astronomical Science, The Graduate University for Advanced Studies (SOKENDAI), 2-21-1 Osawa, Mitaka, Tokyo 1818588, Japan}\email{yukio.katsukawa@nao.ac.jp}		
\author[orcid=0000-0002-0787-8954,sname='Bernasconi']{Pietro~Bernasconi} \affiliation{Johns Hopkins University Applied Physics Laboratory, 11100 Johns Hopkins Road, Laurel, Maryland, USA}\email{pietro.bernasconi@jhuapl.edu}		
\author[sname='Berkefeld']{Thomas~Berkefeld} \affiliation{Institut für Sonnenphysik (KIS), Georges-Köhler-Allee 401a, 79110 Freiburg, Germany}\email{thomas.berkefeld@leibniz-kis.de}		
\author[orcid=0009-0009-4425-599X,sname='Feller']{Alex~Feller} \affiliation{Max-Planck-Institut für Sonnensystemforschung, Justus-von-Liebig-Weg 3, 37077 Göttingen, Germany}\email{feller@mps.mpg.de}		
\author[orcid=0000-0001-6317-4380,sname='Riethmüller']{Tino~L.~Riethmüller} \affiliation{Max-Planck-Institut für Sonnensystemforschung, Justus-von-Liebig-Weg 3, 37077 Göttingen, Germany}\email{riethmueller@mps.mpg.de}		

%%% Sunrise CoIs
%%% Sunrise CoIs		mandatory for all papers
\author[orcid=0000-0001-9228-3412,sname='Álvarez-Herrero']{Alberto~Álvarez-Herrero} \affiliation{Instituto Nacional de T\'ecnica Aeroespacial (INTA), Ctra. de Ajalvir, km. 4, E-28850 Torrejón de Ardoz, Spain}\affiliation{Spanish Space Solar Physics Consortium}\email{alvareza@inta.es}		
\author[orcid=0000-0001-5616-2808,sname='Kubo']{Masahito~Kubo} \affiliation{National Astronomical Observatory of Japan, 2-21-1 Osawa, Mitaka, Tokyo 181-8588, Japan}\email{masahito.kubo@nao.ac.jp}		
\author[orcid=0000-0001-8829-1938,sname='Orozco~Suárez']{David~Orozco~Suárez} \affiliation{Instituto de Astrofísica de Andalucía, CSIC, Glorieta de la Astronomía s/n, 18008 Granada, Spain}\affiliation{Spanish Space Solar Physics Consortium}\email{orozco@iaa.es}		
\author[sname='Grauf']{Bianca~Grauf} \affiliation{Max-Planck-Institut für Sonnensystemforschung, Justus-von-Liebig-Weg 3, 37077 Göttingen, Germany}\email{grauf@mps.mpg.de}		
\author[sname='Carpenter']{Michael~Carpenter} \affiliation{Johns Hopkins University Applied Physics Laboratory, 11100 Johns Hopkins Road, Laurel, Maryland, USA}\email{michael.carpenter@jhuapl.edu}		
\author[sname='Bell']{Alexander~Bell} \affiliation{Institut für Sonnenphysik (KIS), Georges-Köhler-Allee 401a, 79110 Freiburg, Germany}\email{albe@leibniz-kis.de}		
\author[orcid=0000-0001-7764-6895,sname='Martínez~Pillet']{Valentín~Martínez~Pillet} \affiliation{Instituto de Astrofísica de Canarias, Vía Láctea, s/n, E-38205 La Laguna, Spain}\affiliation{Universidad de La Laguna, E-38205 La Laguna, Spain}\affiliation{Spanish Space Solar Physics Consortium}\email{vmpillet@iac.es}
\author[orcid=0000-0001-7696-8665,sname='Gizon']{Laurent~Gizon} \affiliation{Max-Planck-Institut für Sonnensystemforschung, Justus-von-Liebig-Weg 3, 37077 Göttingen, Germany}\affiliation{Institut für Astrophysik und Geophysik, Georg-August-Universität Göttingen, 37077 Gōttingen, Germany}\email{gizon@mps.mpg.de}		%

%%% Early-Career Scientists		mandatory for all papers
\author[orcid=0000-0002-7318-3536,sname='Bailén']{Francisco~Javier~Bailén} \affiliation{Instituto de Astrofísica de Andalucía, CSIC, Glorieta de la Astronomía s/n, 18008 Granada, Spain}\affiliation{Spanish Space Solar Physics Consortium}\email{fbailen@iaa.es}		
\author[orcid=0000-0002-2055-441X,sname='Blanco~Rodríguez']{Julian~Blanco~Rodríguez} \affiliation{Universitat de Valencia, Catedrático José Beltrán 2, E-46980 Paterna, Spain}\affiliation{Spanish Space Solar Physics Consortium}\email{julian.blanco@uv.es}		
\author[orcid=0000-0003-4319-2009,sname='Castellanos~Durán']{Juan~Sebastián~Castellanos~Durán} \affiliation{Max-Planck-Institut für Sonnensystemforschung, Justus-von-Liebig-Weg 3, 37077 Göttingen, Germany}\email{castellanos@mps.mpg.de}		
\author[orcid=0009-0002-6808-5154,sname='Harnes']{Edvarda~Harnes} \affiliation{Max-Planck-Institut für Sonnensystemforschung, Justus-von-Liebig-Weg 3, 37077 Göttingen, Germany}\email{harnes@mps.mpg.de}		
\author[orcid=0000-0001-6029-7529,sname='Hoelken']{Johannes~Hoelken} \affiliation{Max-Planck-Institut für Sonnensystemforschung, Justus-von-Liebig-Weg 3, 37077 Göttingen, Germany}\email{hoelken@mps.mpg.de}		
\author[orcid=0000-0003-1409-1145,sname='Iglesias']{Francisco~A.~Iglesias} \affiliation{Max-Planck-Institut für Sonnensystemforschung, Justus-von-Liebig-Weg 3, 37077 Göttingen, Germany}\affiliation{Grupo de Estudios en Heliofísica de Mendoza, CONICET, Universidad de Mendoza, Boulogne sur Mer 683, 5500 Mendoza, Argentina}\email{iglesias@mps.mpg.de}		
\author[orcid=0000-0002-4669-5376,sname='Ishikawa']{Ryohtaroh~T.~Ishikawa} \affiliation{National Institute for Fusion Science, 322-6 Oroshi-cho, Toki City 509-5292, Japan}\email{ishikawa.ryohtaro@nifs.ac.jp}		
\author[orcid=0000-0001-7452-0656,sname='Kawabata']{Yusuke~Kawabata} \affiliation{National Astronomical Observatory of Japan, 2-21-1 Osawa, Mitaka, Tokyo 181-8588, Japan}\email{kawabata.yusuke@nao.ac.jp}		
\author[orcid=0000-0002-1043-9944,sname='Matsumoto']{Takuma~Matsumoto} \affiliation{Centre for Integrated Data Science, Institute for Space-Earth Environmental Research, Nagoya University, Furocho, Chikusa-ku, Nagoya, Aichi 464-8601, Japan}\email{takuma.matsumoto@gmail.com}		
\author[orcid=0000-0002-7044-6281,sname='Oba']{Takayoshi~Oba} \affiliation{Advanced Research Center for Space Science and Technology, Institute of Science and Engineering, Kanazawa University, Kakuma-machi, Kanazawa, Ishikawa 920-1192, Japan}\affiliation{Max-Planck-Institut für Sonnensystemforschung, Justus-von-Liebig-Weg 3, 37077 Göttingen, Germany}\email{oba@mps.mpg.de}		
\author[orcid=0009-0007-5514-4553,sname='Singh']{Kunal~H.~Singh} \affiliation{Max-Planck-Institut für Sonnensystemforschung, Justus-von-Liebig-Weg 3, 37077 Göttingen, Germany}\email{}		
\author[orcid=0000-0003-0175-6232,sname='Siu-Tapia']{Azaymi~L.~Siu-Tapia} \affiliation{Instituto de Astrofísica de Andalucía, CSIC, Glorieta de la Astronomía s/n, 18008 Granada, Spain}\affiliation{Spanish Space Solar Physics Consortium}\email{siu@iaa.es}		
\author[orcid=0000-0003-1483-4535,sname='Strecker']{Hanna~Strecker} \affiliation{Instituto de Astrofísica de Andalucía, CSIC, Glorieta de la Astronomía s/n, 18008 Granada, Spain}\affiliation{Spanish Space Solar Physics Consortium}\email{streckerh@iaa.es}		
\author[orcid=0000-0003-1971-5551,sname='Vukadinović']{Dušan~Vukadinović} \affiliation{Institut für Physik, Universität Graz, Universitätsplatz 5, 8010 Graz, Austria}\affiliation{Max-Planck-Institut für Sonnensystemforschung, Justus-von-Liebig-Weg 3, 37077 Göttingen, Germany}\email{vukadinovic@mps.mpg.de}		

\author[sname='van~Noort']{Michiel~van~Noort} \affiliation{Max-Planck-Institut für Sonnensystemforschung, Justus-von-Liebig-Weg 3, 37077 Göttingen, Germany}\email{vannoort@mps.mpg.de}		
%	
%%% TuMag CoIs	
\author[orcid=0000-0003-4738-7727,sname='Balaguer~Jiménez']{Maria~Balaguer~Jiménez} \affiliation{Instituto de Astrofísica de Andalucía, CSIC, Glorieta de la Astronomía s/n, 18008 Granada, Spain}\affiliation{Spanish Space Solar Physics Consortium}\email{balaguer@iaa.es}		
\author[orcid=0000-0002-4208-3575,sname='Sanchis~Kilders']{Esteban~Sanchis~Kilders} \affiliation{Universitat de Valencia, Catedrático José Beltrán 2, E-46980 Paterna, Spain}\affiliation{Spanish Space Solar Physics Consortium}\email{esteban.sanchis@uv.es}		
\author[orcid=0000-0001-9272-6439,sname='Torralbo']{Ignacio~Torralbo} \affiliation{Universidad Politécnica de Madrid,  Plaza Cardenal Cisneros 3, E-28040 Madrid, Spain}\affiliation{Spanish Space Solar Physics Consortium}\email{ignacio.torralbo@upm.es}		
\author[orcid=0000-0002-3242-1497,sname='Kuckein']{Christoph~Kuckein} \affiliation{Instituto de Astrofísica de Canarias, Vía Láctea, s/n, E-38205 La Laguna, Spain}\affiliation{Universidad de La Laguna, E-38205 La Laguna, Spain}\affiliation{Spanish Space Solar Physics Consortium}\email{christoph.kuckein@iac.es}%	
%	
%%% SCIP CoIs	
\author[orcid=0000-0001-5686-3081,sname='Hara']{Hirohisa~Hara} \affiliation{National Astronomical Observatory of Japan, 2-21-1 Osawa, Mitaka, Tokyo 181-8588, Japan}\email{hirohisa.hara@nao.ac.jp}		
\author[orcid=0000-0003-4764-6856,sname='Shimizu']{Toshifumi~Shimizu} \affiliation{Department of Earth and Planetary Science, The University of Tokyo, 7-3-1, Hongo, Bunkyo-ku, Tokyo 113-0033, Japan}\affiliation{Institute of Space and Astronautical Science, Japan Aerospace Exploration Agency, 3-1-1, Yoshinodai, Chuo-ku, Sagamihara, Kanagawa 252-5210, Japan}\email{shimizu.toshifumi@isas.jaxa.jp}		
	
\author[sname='Volkmer']{Reiner~Volkmer} \affiliation{Institut für Sonnenphysik (KIS), Georges-Köhler-Allee 401a, 79110 Freiburg, Germany}\email{volkmer@leibniz-kis.de}		
\author[sname='Preis']{Tobias~Preis} \affiliation{Institut für Sonnenphysik (KIS), Georges-Köhler-Allee 401a, 79110 Freiburg, Germany}\email{preis@leibniz-kis.de}		

\author[orcid=0000-0003-2409-3742,sname='Raouafi']{Nour~E.~Raouafi} \affiliation{Johns Hopkins University Applied Physics Laboratory, 11100 Johns Hopkins Road, Laurel, Maryland, USA}\email{Nour.Raouafi@jhuapl.edu}		
\author[orcid=0000-0002-8164-5948,sname='Vourlidas']{Angelos~Vourlidas} \affiliation{Johns Hopkins University Applied Physics Laboratory, 11100 Johns Hopkins Road, Laurel, Maryland, USA}\email{Angelos.Vourlidas@jhuapl.edu}		

%\author[orcid=0000-0002-0787-8954,sname='Bernasconi']{Pietro~Bernasconi} \affiliation{Johns Hopkins University Applied Physics Laboratory, 11100 Johns Hopkins Road, Laurel, Maryland, USA}\email{pietro.bernasconi@jhuapl.edu}	
%\author[sname='Carpenter']{Michael~Carpenter} \affiliation{Johns Hopkins University Applied Physics Laboratory, 11100 Johns Hopkins Road, Laurel, Maryland, USA}\email{michael.carpenter@jhuapl.edu}	
%	
%	
%%% Telescope/ISLiD CoIs	
%\author[orcid=0000-0002-9972-9840,sname='Gandorfer']{Achim~Gandorfer} \affiliation{Max-Planck-Institut für Sonnensystemforschung, Justus-von-Liebig-Weg 3, 37077 Göttingen, Germany}\email{gandorfer@mps.mpg.de}		
%\author[orcid=0000-0001-6317-4380,sname='Riethmüller']{Tino~L.~Riethmüller} \affiliation{Max-Planck-Institut für Sonnensystemforschung, Justus-von-Liebig-Weg 3, 37077 Göttingen, Germany}\email{riethmueller@mps.mpg.de}		
\author[sname='Hirzberger']{Johann~Hirzberger} \affiliation{Max-Planck-Institut für Sonnensystemforschung, Justus-von-Liebig-Weg 3, 37077 Göttingen, Germany}\email{hirzberger@mps.mpg.de}	
\author[sname='Deutsch']{Werner~Deutsch} \affiliation{Max-Planck-Institut für Sonnensystemforschung, Justus-von-Liebig-Weg 3, 37077 Göttingen, Germany}\email{deutschw@mps.mpg.de}		
\author[sname='Germerott']{Dietmar~Germerott} \affiliation{Max-Planck-Institut für Sonnensystemforschung, Justus-von-Liebig-Weg 3, 37077 Göttingen, Germany}\email{germerott@mps.mpg.de}		
\author[sname='Heerlein']{Klaus~Heerlein} \affiliation{Max-Planck-Institut für Sonnensystemforschung, Justus-von-Liebig-Weg 3, 37077 Göttingen, Germany}\email{heerlein@mps.mpg.de}		
\author[sname='Kolleck']{Martin~Kolleck} \affiliation{Max-Planck-Institut für Sonnensystemforschung, Justus-von-Liebig-Weg 3, 37077 Göttingen, Germany}\email{kolleck@mps.mpg.de}		

%%% Flight Operations
\author[orcid=0000-0002-8169-8476,sname='Álvarez~García']{Daniel~Álvarez~García} \affiliation{Instituto de Astrofísica de Andalucía, CSIC, Glorieta de la Astronomía s/n, 18008 Granada, Spain}\affiliation{Spanish Space Solar Physics Consortium}\email{dalvarez@iaa.es}				
\author[orcid=0000-0002-6297-0681,sname='López~Jiménez']{Antonio~C.~López~Jiménez} \affiliation{Instituto de Astrofísica de Andalucía, CSIC, Glorieta de la Astronomía s/n, 18008 Granada, Spain}\affiliation{Spanish Space Solar Physics Consortium}\email{aclopezj@gmail.com}			
\author[orcid=0000-0001-8669-8857,sname='Bellot~Rubio']{Luis~R.~Bellot~Rubio} \affiliation{Instituto de Astrofísica de Andalucía, CSIC, Glorieta de la Astronomía s/n, 18008 Granada, Spain}\affiliation{Spanish Space Solar Physics Consortium}\email{lbellot@iaa.es}		
\author[orcid=0000-0002-5773-0368,sname='Morales-Fernández']{José~Miguel Morales-Fernández} \affiliation{Instituto de Astrofísica de Andalucía, CSIC, Glorieta de la Astronomía s/n, 18008 Granada, Spain}\affiliation{Spanish Space Solar Physics Consortium}\email{jmorales@iaa.es}		
\author[orcid=0009-0002-3396-3359,sname='Moreno~Mantas']{Antonio~Jesús~Moreno~Mantas} \affiliation{Instituto de Astrofísica de Andalucía, CSIC, Glorieta de la Astronomía s/n, 18008 Granada, Spain}\affiliation{Spanish Space Solar Physics Consortium}\email{ammantas@iaa.es}		
\author[orcid=0000-0003-2817-8719,sname='Aparicio~del~Moral']{Beatriz~Aparicio~del~Moral} \affiliation{Instituto de Astrofísica de Andalucía, CSIC, Glorieta de la Astronomía s/n, 18008 Granada, Spain}\affiliation{Spanish Space Solar Physics Consortium}\email{bea@iaa.es}				
\author[orcid=0009-0008-7320-5716,sname='Sánchez~Gómez']{Antonio~Sánchez~Gómez} \affiliation{Instituto de Astrofísica de Andalucía, CSIC, Glorieta de la Astronomía s/n, 18008 Granada, Spain}\affiliation{Spanish Space Solar Physics Consortium}\email{asgomez@iaa.es}				
\author[orcid=0009-0004-3976-2528,sname='Bailón~Martínez']{Eduardo~Bailón~Martínez} \affiliation{Instituto de Astrofísica de Andalucía, CSIC, Glorieta de la Astronomía s/n, 18008 Granada, Spain}\affiliation{Spanish Space Solar Physics Consortium}\email{ebailon@iaa.es}				
\author[orcid=0000-0001-7094-518X,sname='Santamarina~Guerrero']{Pablo~Santamarina~Guerrero} \affiliation{Instituto de Astrofísica de Andalucía, CSIC, Glorieta de la Astronomía s/n, 18008 Granada, Spain}\affiliation{Spanish Space Solar Physics Consortium}\email{pablosantamarinag@gmail.com}				
\author[orcid=0000-0001-5961-1189,sname='Hernández Expósito']{David~Hernández Expósito} \affiliation{Instituto de Astrofísica de Canarias, Vía Láctea, s/n, E-38205 La Laguna, Spain}\affiliation{Spanish Space Solar Physics Consortium}\email{david.hernandez.exposito@iac.es}		
\author[orcid=0009-0009-4178-4554,sname='Tobaruela']{Angel~Tobaruela} \affiliation{Instituto de Astrofísica de Andalucía, CSIC, Glorieta de la Astronomía s/n, 18008 Granada, Spain}\affiliation{Spanish Space Solar Physics Consortium}\email{angelt@iaa.es}				
\author[orcid=0000-0002-1225-4177,sname='Gasent~Blesa']{José~Luis~Gasent~Blesa} \affiliation{Universitat de Valencia, Catedrático José Beltrán 2, E-46980 Paterna, Spain}\affiliation{Spanish Space Solar Physics Consortium}\email{jose.l.gasent@uv.es}				
\author[sname='Schulze']{Erich~Schulze} \affiliation{Johns Hopkins University Applied Physics Laboratory, 11100 Johns Hopkins Road, Laurel, Maryland, USA}\email{erich.schulze@jhuapl.edu}				
\author[sname='Eaton']{Harry~Eaton} \affiliation{Johns Hopkins University Applied Physics Laboratory, 11100 Johns Hopkins Road, Laurel, Maryland, USA}\email{harry.eaton@jhuapl.edu}				
\author[sname='Palo']{Geoffrey~Palo} \affiliation{Johns Hopkins University Applied Physics Laboratory, 11100 Johns Hopkins Road, Laurel, Maryland, USA}\email{geoffrey.palo@jhuapl.edu}				
\author[orcid=0009-0006-7164-0576,sname='Ayoub']{Daniel~Ayoub} \affiliation{Johns Hopkins University Applied Physics Laboratory, 11100 Johns Hopkins Road, Laurel, Maryland, USA}\email{Daniel.Ayoub@jhuapl.edu}	
\author[orcid=0000-0001-6793-8528,sname='Naito']{Yoshihiro~Naito} \affiliation{Department of Astronomical Science, The Graduate University for Advanced Studies (SOKENDAI), 2-21-1 Osawa, Mitaka, Tokyo 1818588, Japan}\affiliation{National Astronomical Observatory of Japan, 2-21-1 Osawa, Mitaka, Tokyo 181-8588, Japan}\email{yoshihiro.naito@grad.nao.ac.jp}			
\author[orcid=0000-0001-5518-8782,sname='Quintero Noda']{Carlos~Quintero Noda} \affiliation{Instituto de Astrofísica de Canarias, Vía Láctea, s/n, E-38205 La Laguna, Spain}\affiliation{Universidad de La Laguna, E-38205 La Laguna, Spain}\affiliation{Spanish Space Solar Physics Consortium}\email{carlos.quintero@iac.es}		
\author[sname='Uraguchi']{Fumihiro~Uraguchi} \affiliation{National Astronomical Observatory of Japan, 2-21-1 Osawa, Mitaka, Tokyo 181-8588, Japan}\email{fumihiro.uraguchi@nao.ac.jp}				
\author[orcid=0000-0002-8342-8314,sname='Tsuzuki']{Toshihiro~Tsuzuki} \affiliation{National Astronomical Observatory of Japan, 2-21-1 Osawa, Mitaka, Tokyo 181-8588, Japan}\email{toshihiro.tsuzuki@nao.ac.jp}			
\author[orcid=0000-0002-3787-9640,sname='Piqueras~Carreño']{Javier~Piqueras~Carreño} \affiliation{Universidad Politécnica de Madrid,  Plaza Cardenal Cisneros 3, E-28040 Madrid, Spain}\affiliation{Spanish Space Solar Physics Consortium}\email{javier.piqueras@upm.es}

%-------------------------
%\input{abstract}
%!TEX root = main.tex
%------------------------------------------
\begin{abstract}
	\sunriseiii\ is a stratospheric balloon-borne solar observatory with a 1-m diameter telescope and three post-focus instruments, along with an image stabilisation system, all within a protective gondola. It samples the lower solar atmosphere, from the solar surface to the middle chromosphere, at a resolution approaching 50~km on the Sun. \sunriseiii\ flew successfully for 6.5 days suspended from a zero-pressure stratospheric balloon from northern Sweden to north-western Canada in July 2024, gathering around 200 TB of data. The present issue of ApJL focuses on the first scientific results from the data collected during that flight. 
	This paper introduces this Focus Issue, providing a very brief overview of the capabilities of the instrumentation, the flight and of the gathered data. Challenges for the measurements, data reduction and interpretation are also briefly touched upon. 
	The paper ends with an overview of the first set of science results obtained from these data, as presented in the current  Focus Issue.
\end{abstract}

\keywords{}
%--------------------------
%\input{introduction}
%!TEX root = main.tex
\section{Introduction} 
\label{sec:intro}
Magnetoconvection, i.e., the interaction between magnetic fields and (turbulent) convection, at and below the solar surface is responsible for structuring and energizing the solar atmosphere, in particular the chromosphere and corona \citep[{e.g., }][]{klimchuk2006, wedemeyeretal2009, stein2012}. 
Although the interaction takes place on many spatial scales, of particular interest are the smallest ones accessible to observations, where much of the coupling occurs.

A major challenge facing observational studies of solar small-scale processes from the ground is the variable turbulence inherent to the Earth’s atmosphere. The seeing produced by this turbulence leads to image distortion and smearing. Although it can be compensated using adaptive optics \citep{Rimmele-Marino2011} combined with advanced restoration techniques such as Multi-Object Multi-Frame Blind Deconvolution \citep[MOMFBD;][]{vannoortetal2005} to produce remarkably sharp, diffraction-limited images and partly also polarised spectra \citep{vannoortetal2025}, 
%\todo{ REF TO DKIST ARTICLE} 
the resulting diffraction-limited time series are often of short duration, and any narrow-band data suffer from relatively high noise. 

High-resolution solar missions flying above (most of) the Earth’s atmosphere can produce long time series or large, low-noise and low-straylight spectro-polarimetric scans at close to the telescope’s diffraction limit. The latter is particularly important when determining the magnetic vector, due to the often low signal levels in the linearly polarised Stokes parameters. 
A pioneering space mission aiming to uncover the secrets of magnetoconvection and its effects was the Hinode spacecraft \citep{kosugietal2007}, 
which carried the 50\,cm Solar Optical Telescope \citep[SOT;][]{tsunetaetal2008} and reached a resolution of 0.2\arcsec{}--0.3\arcsec{}. Already such a small telescope by today’s standards still led to many remarkable discoveries and advances \citep[see][]{hinodereviewteam2019}.

To achieve a higher spatial resolution, a larger telescope is needed. So far, solar telescopes larger than the SOT have been carried aloft only by stratospheric balloons. At float altitude, these fly above roughly 99\%\ of the Earth’s atmosphere, so that seeing ceases to be a problem, as long as the Sun is sufficiently far above the horizon (which is typically the case for an elevation above $0\deg$ at float altitude). The first high-resolution balloon-borne solar observatory with a primary mirror diameter of 50\,cm or more\footnote{There was one flight carrying a telescope with a 1\,m mirror} was the Soviet Stratospheric Solar Observatory, which reached a spatial resolution of 0.2\arcsec{} \citep{krat1972,krat1977,krat1981}. Later, the 80\,cm aperture telescope of Flare Genesis \citep[][]{murphyetal1996, bernasconietal1999}, which flew successfully in 1996 and 2000. It provided observations with a resolution  of 0.5\arcsec, which illustrates the challenges that high resolution balloon payloads have to face. These observations enabled new insights into Ellerman bombs and the underlying magnetic structure \citep{georgoulisetal2002,bernasconietal2002}. 

Flare Genesis was followed by Sunrise \citep{bartholetal2011, gandorferetal2011, berkefeldetal2011}, composed of a 1\,m diameter telescope and originally two instruments, a UV imager \citep[SUFI;][]{gandorferetal2011} and an imaging magnetograph \citep[IMaX;][]{martinezpilletetal2011}. Sunrise flew successfully in 2009 \citep{solankietal2010} and 2013 \citep{solankietal2017} producing data that reached a resolution of about 0.075\arcsec\ (by SuFi in 2013). %\hns{\citep{gandorferetal2011} this one?}. Looking at it again, I feel that we don't have to give a reference here.
These flights are now referred to as Sunrise~{\sc i} and {\sc ii}. They produced a string of results that were published in two Focus Issues in 2010 and 2017, containing 12 and 18 papers, respectively. These were later complemented by individual publications, so that the total number of refereed papers from the first two flights now exceeds 100. Many exciting results were obtained. These include, i) the first resolved observations of kG magnetic elements in the quiet internetwork of the Sun \citep{laggetal2010}, ii) the highest observed contrasts of granulation \citep{hirzbergeretal2010} and bright points \citep{riethmulleretal2010} till then, iii) the varied dynamics of bright points \citep{jafarzadehetal2013}, iv) the evolution of ubiquitous linear polarisation patches \citep{danilovicetal2010a}, v) the discovery of horizontal granular vortices \citep{steineretal2010} and the exploration of vertical convective vortices \citep{bonetetal2010}, vi) the discovery of a new mechanism to heat the corona \citep{chittaetal2017}, vii) the discovery of an order of magnitude higher flux emergence rate in the quiet Sun than previously observed \citep{smithaetal2017}, and many more.

After the flight in 2013 the telescope was refurbished while nearly every other part of the observatory was either exchanged or upgraded. Thus the observatory was furnished with three new instruments, a new light-distribution unit, a new gondola and far more powerful instrument control unit and data storage system. This new observatory, referred to as \sunriseiii, flew successfully in July 2024 (after an aborted  flight in 2022), producing a plethora of high-quality science data. The observatory (shown hanging from its launch vehicle in Fig.~1) and its instruments are described in a topical collection of the journal Solar Physics composed of seven publications \citep{korpi-laggetal2025, bernasconietal2025, felleretal2025, iglesiasetal2025, deltoroiniestaetal2025, berkefeldetal2026,katsukawaetal2026}. The present Focus Issue of the Astrophysical Journal Letters presents the first scientific results of \sunriseiii.  

\begin{figure}[t]
	\centering
	\includegraphics[width=\hsize]{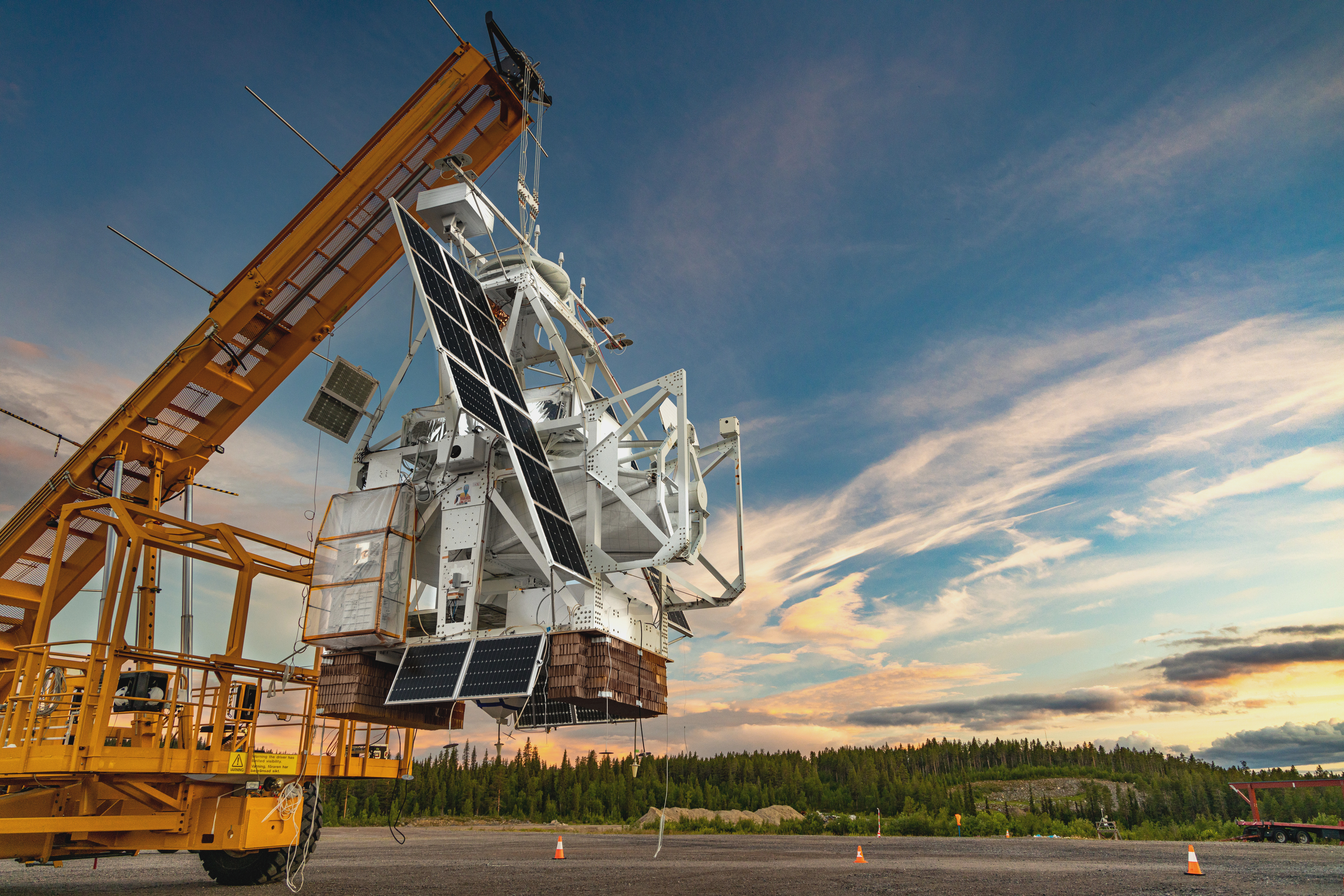} 
	\includegraphics[width=0.8\hsize]{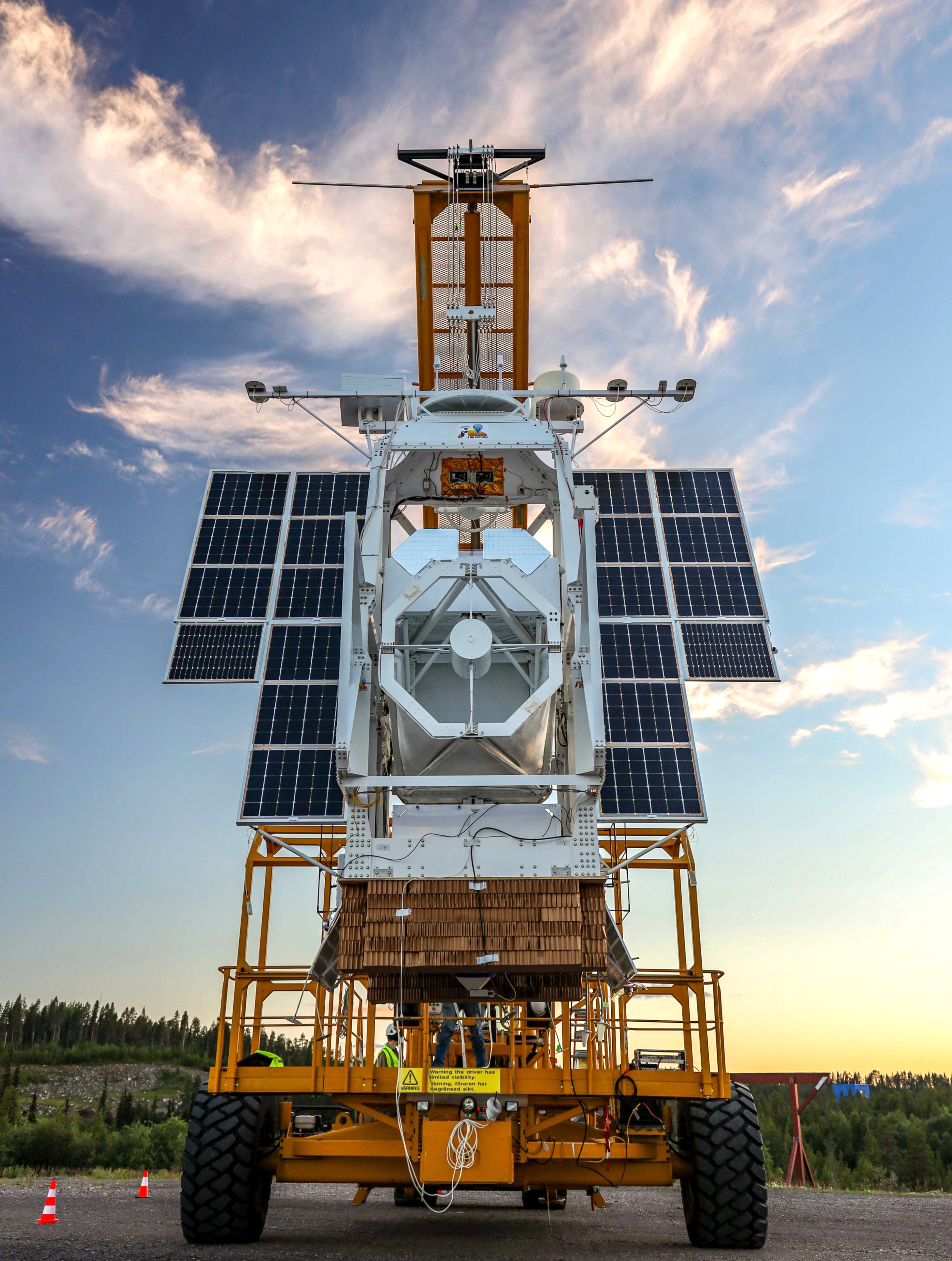}
	\includegraphics[width=\hsize]{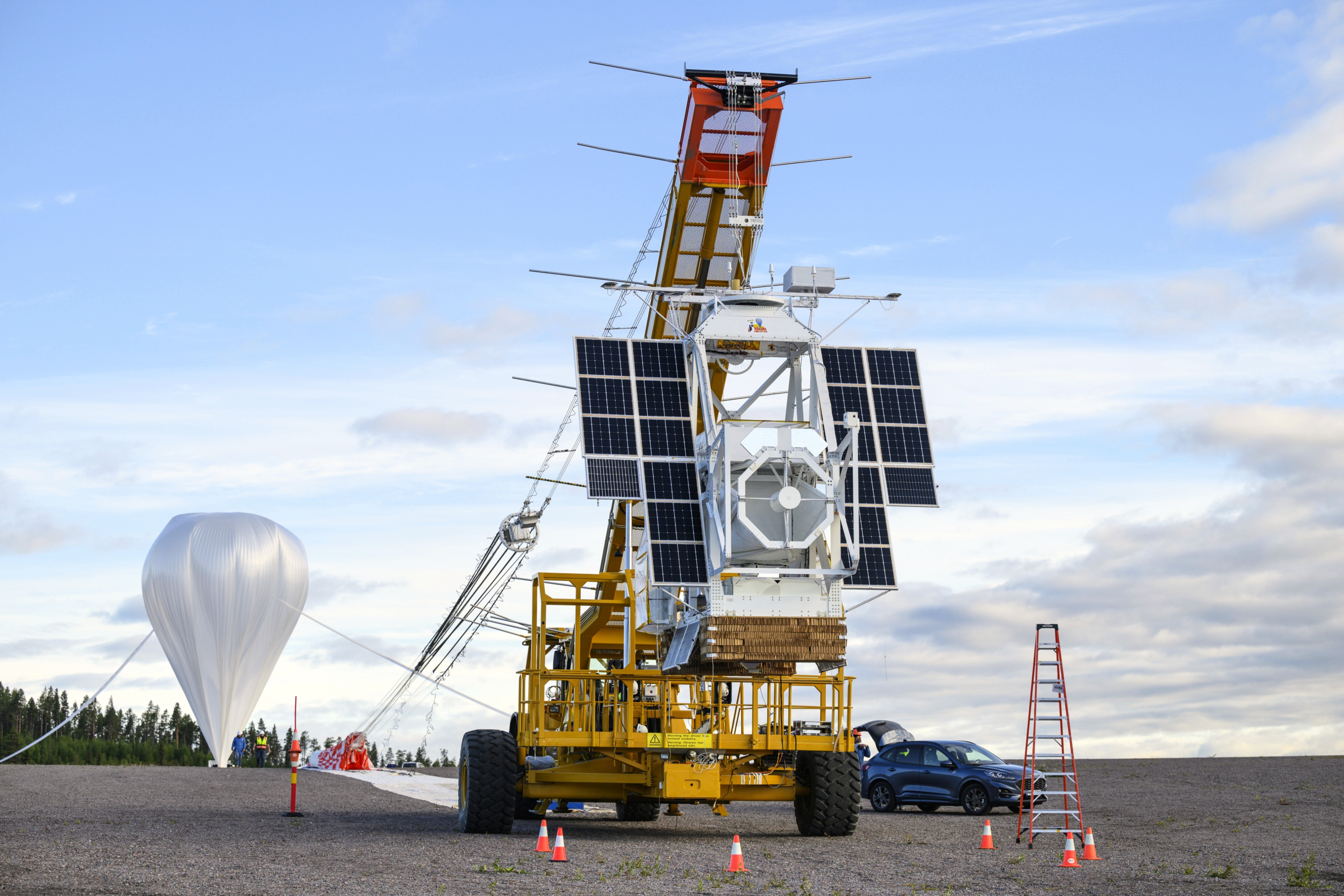}
	\caption{Three images of the fully assembled \sunriseiii\ observatory hanging from the Hercules launch vehicle at the European Space and Sounding Rocket Range (ESRANGE) near Kiruna in northern Sweden prior to launch. The top and middle panels show photographs taken in 2022, the bottom panel shows one from 2024.}
	\label{fig:1000}
\end{figure}
%\todo{FIGURE 1: Three images of Sunrise hanging }

%\input{science_capabilities}
%!TEX root = main.tex
\section{Science capabilities of \sunriseiii}

\sunriseiii\ is a versatile solar observatory probing the Sun in spectral bands reaching from the near ultraviolet (UV, from 309\,nm) to the near infrared (IR, 855\,nm). These wavelengths are well accessible from its float altitude of 35-37 km and allow it to cover the entire height range of the solar photosphere and a large part of the chromosphere. 

The main goal of \sunriseiii\ is to investigate the structure, magnetism, dynamics and energetics of the photosphere and chromosphere at high spatial resolution. This includes probing the complex physics of the interaction between convection, magnetism and oscillations (magnetoconvection) in the photosphere over a wide range of magnetic flux and in different regimes of plasma beta, $\beta = 8\pi p / B^2$ in cgs units. Here $p$ is the gas pressure and $B$ is the magnetic field strength. These regimes vary from the quiet Sun, via network and facular regions to pores and sunspots. \sunriseiii\ also probes the myriad dynamic and thermal effects that this interaction produces in the photosphere and the chromosphere. 

\sunriseiii\ science mainly concentrates on uncovering fine-scale effects, especially those profiting from spectro-polarimetric information, from time series recorded under constant observing conditions and from co-spatial observations covering a broad wavelength range (possible because of negligible differential atmospheric refraction). Another advantage is access to wavelengths not reachable from the ground such as the near UV and e.g. the \ion{K}{1} D$_1$ and D$_2$ lines around {770 nm}, one of which is strongly blended with telluric {O$_2$} lines when observed from the ground. Further strengths are the almost complete absence of the day-night cycle for observations made at high latitudes in summer as well as the near-absence of atmospheric stray light and of differential atmospheric refraction between different wavelengths.

Within the above overarching goals, \sunriseiii\ has a long list of specific scientific questions and topics that can be addressed with the data recorded during its science flight in 2024. Some of the main ones are listed below. 
\begin{enumerate}
	\item {\it Quiet Sun magnetism:} The quiet Sun is an inhomogeneous region composed of strong-field network regions and the internetwork permeated by  weak and often heavily inclined fields (whose origin is different in the photosphere and the chromosphere). See {e.g., }\cite{solanki1993, dewijnetal2009, bellotrubio2019, rempeletal2023} for reviews. 
	\sunriseiii\ has recorded time series, spectral scans and sit-and-stare spectral series of quiet Sun regions. These allow probing the structure and evolution of magnetic features from the low photosphere to the upper chromosphere in both the network and the internetwork regions. The observatory can follow their interactions as well as the waves and oscillations that they harbour through the photosphere and chromosphere. 
	\item {\it Small-scale turbulent dynamo (SSD):} Long predicted by MHD simulations \citep[e.g.,][]{voegler2007,rempel2014}, the definite observational proof of a solar SSD has not yet been achieved, although there have been important contributions towards its identification \citep[e.g.][]{danilovicetal2010b,buehleretal2013,litesetal2014}. The high sensitivity of \sunriseiii\ data to weak fields, the high spatial resolution and the long timeseries will allow setting new constraints on the presence and properties of an SSD acting near the solar surface.
	\item {\it Structure, evolution and dynamics of sunspots:} Sunspots are perhaps the best-known solar magnetic features, lying at the heart of solar activity and variability \citep[{e.g., }][]{solanki2003, Rempel-schliche2011,borrero-ichimoto2011} \sunriseiii\ data can help deduce the physical structure and evolution of components of sunspots such as umbral dots, penumbral filaments, light bridges, etc. as well as enabling the study of the rich spectrum of oscillations and waves found in sunspots. 
	\item {\it Magnetic flux emergence:} Magnetic flux emergence is a complex and poorly understood phenomenon \citep[{e.g., }][]{cheung-isobe2014, weberetal2023}. \sunriseiii\ has recorded the emergence of magnetic flux and the photospheric and chromospheric activity it produces in an active region and in the quiet Sun at high resolution, over a large height range and over a significant period of time. This allows, among other topics, the physics of Ellerman bombs and of arch-filament systems to be investigated in great detail. 
	\item {\it Faculae and plage:} A significant amount of the magnetic flux in solar active regions appears at the surface in the form of faculae and is visible in the chromosphere as plage \citep[e.g.,][]{solankietal2006}. These magnetic features lie at the heart of considerable magnetic activity as the footpoints of active region coronal loops. The stratification, spatial properties and the evolution of solar faculae and the corresponding chromospheric plage features can be ideally addressed by \sunriseiii\ spectro-polarimetric observations. 
	\item {\it Flares:} The mechanisms leading to flaring and the fine structures presented by the flaring plasma are still under debate \citep[e.g.,][]{priest-forbes2002, shibata-magara2011, fletcheretal2011}. High resolution observations of flares, such as those recorded by \sunriseiii{}, can help answer questions about the role of the lower atmosphere in flares including the fine structure of, e.g., flare ribbons. 
	\item {\it Spicules and related chromospheric jets:} Spicules and their relatives are thought to be an important source of energy to the solar chromosphere and possibly corona and may provide some of the mass of the solar wind \citep[{e.g.,}][]{depontieuetal2007, tsiropoulaetal2012}. They are thought to be produced by the interaction of magnetic features in the solar photosphere \citep{samantaetal2019}. New knowledge and understanding of these features will profit not just from the constant high spatial resolution photospheric and chromospheric spectro-polarimetry, but also from the low noise and the very low stray light reached by \sunriseiii\ data. 
	\item {\it Near UV solar polarised spectrum:} Although atlases of the solar spectrum in the near UV exist 
	\citep[{e.g.,}][]{kuruczetal1984,kurucz2005,neckel1984, gandorfer2005}, they are limited to observations of the spatially averaged quiet Sun. \sunriseiii\ provides high spatial resolution scans in polarised spectra covering the wavelength range from 309 to 417 nm. Such observations will allow a wide range of solar features to be studied in this spectral range. The changes in line profiles in different solar features could help to check atomic parameters and possibly improve them \citep{vukadinovicetal2024}.

	\item {\it Excitation, propagation and dissipation of waves:} Magnetically excited and guided waves are observed everywhere in the solar atmosphere and are prime candidates for transporting energy to heat the solar chromosphere and corona \citep[{e.g.,}][]{nakariakov-verwichte2005, jessetal2015}. The (high-cadence) time series recorded under constant high spatial resolution conditions by \sunriseiii\ are ideal for studies of waves and oscillations in the solar photosphere and chromosphere. The fact that many spectral lines with different formation heights and properties are observed simultaneously also opens up the  possibility of following the propagation of waves with height.
	
	\item {\it Removal of magnetic flux from the solar photosphere:} Although there are many studies of the emergence or appearance of new magnetic flux on the solar surface, its removal is far less well studied and is often parameterized in models of surface magnetic flux transport \citep{Jiangetal2014}. To probe the removal of magnetic flux from the solar surface, sensitive, high resolution spectro-polarimetric time series of constant quality are required. \sunriseiii\ provides such data covering multiple heights. 
	
	\item {\it Solar spectral irradiance:} The contrast of magnetic features in the near-UV plays a critical role in modelling the amplitude of solar spectral irradiance variations of relevance for global climate change. This is partly because of the large variability in the UV, but also because this radiation has a direct influence on stratospheric chemistry %\sks{Haigh JD. 2007. LRSP 4, 2} 
	\citep[{e.g.,}][]{haigh2007, solankietal2013}. The high resolution observations in the UV spectral range taken by \sunriseiii\ at different positions on the disc allow characterizing the sources of solar UV irradiance. 
	\item {\it Centre-to-limb-variations:} Observing features such as the quiet Sun, faculae and sunspots at different positions on the solar disc is of interest not just to solar physicists, but also for the stellar and indirectly the exoplanetary community. Thus granular and facular/plage contrasts and the magnetic influence on line asymmetries and line shifts need to be known on different parts of the solar disc to be applicable to other, spatially unresolved stars. By observing different parts of the solar disc, \sunriseiii, can contribute to such studies, as well as to the investigation of solar irradiance variations.  
\end{enumerate}

%\input{instrumentation}
%!TEX root = main.tex
\section{Instrumental capabilities and challenges}
\sunriseiii\ is composed of a Gregory telescope with a 1 m diameter main mirror, an active secondary and two deflecting mirrors that feed light into the Post-Focus Instrumentation (PFI) unit located above the telescope. Inside this compartment there is an Image Stabilisation and Light Distribution (ISLiD) unit that distributes the radiation to the three science instruments and the Correlating Wavefront Sensor (CWS) while ensuring that each science instrument gets all the light at its operating wavelengths. The ISLiD unit also contains a rapid tip-tilt mirror controlled by the CWS, which is responsible for the fine stabilisation of the image beyond what the gondola can achieve (see below), and for the autofocus and coma measurement. 

The science instruments carried by \sunriseiii\ are composed of two slit spectro-polarimeters (SCIP and SUSI) and a multi-channel vector magnetograph (TuMag). These, along with the CWS and the gondola are very briefly introduced below. Here we concentrate on their properties relevant for the observations. For technical details, the reader is referred to the instrument papers cited below.

\begin{itemize}
	\item {\it SCIP, the Sunrise Chromospheric Infrared Spectro-Polarimeter:} 
	SCIP \citep{katsukawaetal2026} is a scanning slit (Littrow) spectro-polarimeter with fixed high-order echelle grating. It measures the full Stokes vector simultaneously over two wavelength ranges (765-771~nm and 846-855~nm), which include a number of photospheric (e.g., \ion{Fe}{1} 846.8~nm with a Land\'{e} factor of $g=2.5$) and chromospheric lines (mainly two of the \ion{Ca}{2}\,IR triplet lines, 854.2~nm (effective Land\'e factor $g_{\rm eff}=1.10$) and 849.8~nm ($g_{\rm eff}=1.07$) as well as the \ion{K}{1} lines at 766.4~nm ($g_{\rm eff}=1.16$) and 769.8~nm ($g_{\rm eff}=1.33$)).  Spectral resolving power is $\lambda/\Delta\lambda = 2 \times 10^5$.
	
	The field-of-view (FOV) depends on the length of the scan, from no scan (sit-and-stare mode) to a maximum FOV of $58\arcsec\times58\arcsec$, where the length of the slit is 58\arcsec\ and 58\arcsec\ is the maximum scan width.  SCIP has different scan modes. Standard measurements record $I,Q,U,V$ in dual beam and differ from each other in terms of scan length and integration time, which is a maximum of 10~s per slit position (which results to a noise level of 0.03\%\ of the continuum intensity for measurements carried out at disc centre). By using synchronized, continuous polarization modulation and onboard demodulation to minimize integration dead time, a 10 s integration can achieve polarization sensitivities of 0.04\%\ ($1\sigma$) and 0.03\%\ ($1\sigma$) in the 850\,nm and 770\,nm channels, respectively.  The integration time determines the scan speed. At 1~s (10~s) integration time per slit position, it takes 12~min (107 min) to scan the full FOV (621 scan steps in all, with a step size of 0.094\arcsec).
	
	SCIP also has a rapid mode in which only Stokes $I$ is measured (allowing much faster scan speeds). In this mode, the maximum FOV can be scanned in 40 sec. 
	SCIP also has a slit-jaw (SJ) imager operating in the continuum at 770.5 nm with a field of view of $60\arcsec\times 60$\arcsec.
	
	\item{\it SUSI, the \sunriseiii\ Ultraviolet Spectropolarimeter and Imager:}
	SUSI \citep{felleretal2025} is a scanning slit (modified Czerny-Turner) full-Stokes, dual-beam spectro-polarimeter with tunable low-order grating. It observes in a 1.7--2.7\,nm wide wavelength range within the limits of 309.0\,nm and 417.0\,nm. The most prominent spectral lines in the SUSI wavelength range are the chromospheric \ion{Ca}{2}~H and K lines. Among the many other, often poorly studied lines in this wavelength range are over 150 formed at least partly in the chromosphere. See, e.g., \cite{riethmueller2019} for a discussion of two of the wavelength ranges that SUSI observes, as well as \cite{harnesetal2025}. SUSI can resolve regions as small as 56 km at the shortest and 76 km at the longest wavelength it observes in. Its 60\arcsec\ long slit can scan a FOV of 58\arcsec. The spectral sampling varies between 8.5 and 13.5\,m\AA/px (0.85–1.35\,pm/px), depending on grating diffraction order and wavelength. 
	%The SUSI polarization measurement is based on a zero-order rotating waveplate, resulting in four full-Stokes measurements every 1.024\,s. 
	The polarimetric sensitivity of SUSI changes within the covered NUV spectral range, as illustrated in Figure\,16 of \cite{felleretal2025}.  E.g., after 1\,s integration it reaches $\approx2\%$ and $\approx1\%$ at 328\,nm and 400\,nm, respectively, for linear polarization. 
	With the first version of the calibration pipeline, and further spatio-temporal integration, SUSI data reaches a sensitivity $\approx0.5\%$. Further improvements to reach the target of $0.1\%$ are under development.

	Like SCIP, SUSI also has different scan modes and a SJ imager that images the Sun over a $22\arcsec\times 60\arcsec$ field-of-view. The SJ imager sees the Sun through a fixed band-pass filter centered on 325.4 nm, and with a FWHM of 0.9 nm. It simultaneously records a focused and a defocused image, allowing the SJ images as well as the spectral scans to be numerically reconstructed post-facto with the help of this phase diversity (PD) information.
	
	\item{\it TuMag, the Tunable Magnetograph:} 
	TuMag \citep{deltoroiniestaetal2025} is an imaging dual-beam vector magnetograph that can sequentially observe two spectral lines out of the following three, the photospheric \ion{Fe}{1} 525.02\,nm (Land\'{e} factor, $g=3$), \ion{Fe}{1} 525.06\,nm  (effective Land\'{e} factor, $g_{\rm eff}=1.5$), and the \ion{Mg}{1} b$_2$ line at 517.27\,nm  ($g_{\rm eff}=1.75$) line whose core is formed in the lower chromosphere. The FOV is fixed at $63\arcsec \times 63\arcsec$. The spectral resolution is 8.7~pm (defined here as the FWHM of the spectral transmission profile). 
	
	TuMag has a number of observing modes that differ according to the number of spectral lines recorded, the number of wavelength samples per line (ranging between 3 and 12) and whether only Stokes $I$ is recorded, or $I$ and $V$, or the full Stokes vector. The cadence ranges from 30 to 90~s, depending on the observing mode. TuMag also has PD capability, allowing images to be reconstructed to overcome optical aberrations.
\end{itemize}

In summary, each of the three instruments covers both the photosphere and at least parts of the chromosphere. Together, they sample these atmospheric layers almost as completely as can be done by observations limited to the spectral range available to \sunriseiii.  The three instruments also complement each other in a number of ways, most obviously by the different choices made when trading off the combination of instantaneous imaging together with spectral scanning (TuMag) vs. instantaneous spectral coverage coupled with spatial scanning (SCIP and SUSI). In addition, SCIP concentrates on 2 fixed spectral ranges that are always recorded together, while SUSI can only record one out of a larger number of possible wavelength ranges at a time. Each instrument also observes the Sun in a different wavelength band. 

Importantly, the SCIP and SUSI slits lie parallel to each other and scan in the same direction. For a number of observations they were coaligned over the whole scan. Thanks to the very thin atmosphere at float altitude, wavelength dependent differential refraction effects in the Earth’s atmosphere are negligible, so that SCIP and SUSI spectra largely sensed the same parts of the Sun.

The 1 m aperture main telescope of Sunrise allows resolving solar features at below 60 km at SUSI's shortest wavelength. Although this is not as high as the spatial resolution the Daniel K. Inouye Telescope \citep[DKIST;][]{rimmeleetal2020} can achieve \citep[See][]{Kuridzeetal2025}, Sunrise has unique advantages. The absence of seeing at float altitude (as long as the Sun was higher than $0^\circ$ elevation) allowed extended nearly undistorted time series to be recorded at close to the diffraction limit. More importantly, accurate and low-noise polarimetric observations could be carried out regularly during the flight, without any danger of seeing-induced cross-talk (although jitter was an issue --- see below). This capability is of particular value for magnetic field measurements in the quiet Sun and in the chromosphere, due to the low signals present there. 
Another strength of the observatory is the low spatial scattered light,  allowing emission lines to be recorded just outside the disc even in the blue part of the spectrum.

	Attempting diffraction limited observations from a balloon payload is not without challenges. These include obtaining a stable image without jitter induced by wind, rotation of the balloon and oscillations of the balloon-payload system (which together form a nearly 300 m long pendulum), avoiding image distortions due to gravitational deformation of the lightweighted main mirror, or thermal changes, e.g., due to changing solar elevation, etc. Although the Sun never set on \sunriseiii\ at float altitude, due to the unusually southerly trajectory, the Sun dipped below an elevation of zero degrees each night, leading to strong seeing as the air mass of the Earth's atmosphere increased significantly for about 2.5--4.5 hours on all 6 nights. 
	
	The biggest challenge is ensuring stable pointing while suspended from a balloon that is at the mercy of the prevailing winds. To achieve the milli-arcsecond stability required for diffraction-limited images, a combination of the stabilized pointing provided by the gondola and the rapid motions of the tip-tilt mirror guided by the CWS was used. The gondola used a full-Sun guide telescope with a quad cell to keep the solar disc aligned. The information from this telescope (which was offset relative to the main telescope to allow the latter to observe different parts of the solar disc) and from an Inertial Measurement Unit (IMU) was fed into a computer that then drove three sets of motors controlling azimuth, elevation and roll of the combined gondola-telescope system. 
	The residual $1\,\sigma$ jitter in flight (averaged over 62~hrs of science data acquisition) was 3.2\arcsec\ in azimuth and 1.9\arcsec\ in elevation. Further details on the gondola and its performance can be found in \citet{bernasconietal2025}.
	
	To enhance the image stability further to the levels needed for diffraction limited observations, the CWS was employed. It separately compensates for slowly varying low-order optical aberrations and more rapid jitter. For the slow compensation of aberrations such as defocus and coma, it employs a Shack-Hartmann wavefront sensor. The actual correction is carried out by the active secondary mirror of the telescope. For the image stabilisation (jitter reduction), a correlation tracker that can lock on granulation even close to the limb, controls a rapid tip-tilt mirror in the light path. The CWS can reduce jitter at frequencies below 150~Hz (its sensor operates at 7~kHz). The residual jitter while observations were being made during the 2024 flight was better than 0.01\arcsec\ in one direction, but closer to 0.02\arcsec\ in the other. The small FoV of 6.7\arcsec of the CT of the CWS was offset by 15\arcsec with respect to the centre of the TuMag FoV to allow CWS to lock onto solar on-disc features while doing off-limb observations. Technical details and performance during the flight are thoroughly described by \cite{berkefeldetal2026}.

	Although the gondola and the CWS performed exceedingly well, some of the data were affected by residual jitter, whose magnitude varied over the course of the flight. Also, occasionally, the CWS lost its lock on the solar feature it was tracking, so that the image jumped, interrupting time series and/or spatial scans by the spectrographs.

	Other points of importance for data analysis are that the solar scene rotates over the day at a non-constant rate, as shown in Fig.~7 of \cite{korpi-laggetal2025}. Note that the rotation is centred on the centre of the CWS/CT FoV. This is of particular importance for the SCIP and SUSI slits, which follow a curved trajectory when scanning the solar surface and for which sit-and-stare observations or very short, repeated scans cover the same parts of the solar surface only in the central part of the slit if repeated for a sufficiently long time. For this reason such observations were preferentially placed at times of low image rotation (local mornings and evenings). 
	
	The fields of view of the instruments were carefully aligned on ground and during flight, such that the spectrograph scans always stayed within the FoV covered by TuMag. Scans were typically centred on the middle of the TuMag FoV, with the exception of very narrow, repeated scans, which were centred on the CWS/CT FoV, i.e. on the centre of image rotation \citep[see Fig.~8 of][]{korpi-laggetal2025}.

%\input{observations}
%!TEX root = main.tex
\section{Observed solar objects and overview of collected data}
During the roughly 6.5 day-long flight, \sunriseiii\ pointed at the Sun for 152.2 h, maintaining lock on solar targets for 102.7 h. The remaining time was mainly used for calibration measurements (finding solar limbs to accurately determine the position on the solar disc, dark current measurements, flat fields, PD measurements in the case of TuMag), or to select and lock on the correct solar target (and wavelength region in the case of SUSI). Some observing time was lost because of seeing when the Sun was partly behind the Earth’s atmosphere (although a large fraction of this time was used for calibration purposes), and a few hours were spent recovering from errors.

Careful planning maximised the use of available observation time.  A timeline concept was developed before the flight. This optimised the synchronisation of the three science instruments and pointing system. 
The timeline comprised individual observing blocks. Each block consisted of a sequence of commands defining observing modes and identifying necessary calibration measurements.  During daily science planning meetings, these up to eight-hour blocks were scheduled to run on manually selected solar targets.

%Careful planning 
This timeline concept 
allowed most of the originally planned observations to be carried out, with the 240 TB hard discs being 80\% full by the end of the mission. 
%The data are public and can be downloaded from …. 
The list of targets includes the quiet Sun, sunspots, emerging flux regions, plage, flares, spicules at the poles, the East and West limbs, 
a coronal hole, filaments and an arch-filament system. Nearly 31\% of the science observations were dedicated to the quiet Sun, followed by sunspots (22\%) and plage (12\%) while other targets were observed less often. Although \sunriseiii\ concentrated on regions close to disc centre, it did also observe regions close to and at the limb as well as at intermediate $\mu=\cos\theta$, where $\theta$ is the heliocentric angle.

The new state-of-the art gondola in combination with the CWS ensured constant pointing and feature tracking over extended periods of time, making it possible to acquire several long time-series datasets, including an uninterrupted four hours long observation of an emerging flux region. On 13th July 2024, \sunriseiii\ successfully caught an M5.3 flare, and one hour of pre-flaring activity in the active region AR13738. These observations are three hours long and were carefully planned using the $k$-scheme flare prediction model \citep{kusanoetal2020} run some hours earlier.

All the science data were stored onboard. Typically only selected thumbnails were transferred to the ground. Only very occasionally could full images be transferred to check the data quality. 

An overview of all the observations recorded by \sunriseiii\ is presented in Table~\ref{tab:table}. 
Note that SCIP observed all the targets listed in the Table. It is not explicitly listed in Table~\ref{tab:table} because it always recorded the same standard spectral ranges.

\begin{deluxetable*}{lllllll}
	\tablecaption{An overview of all the observations recorded by \sunriseiii.  \label{tab:table}.}
	\tablehead{
		\multicolumn{1}{l}{Target} & \multicolumn{1}{l}{Sunrise$\_$ID} & \multicolumn{1}{l}{AR No.} & \multicolumn{1}{l}{Observing time} & \multicolumn{1}{l}{$\mu$} & \multicolumn{2}{c}{Spectral region$^1$}\\
		\colhead{ } & \colhead{ } & \colhead{ } & \colhead{ } & \colhead{ } & \multicolumn{1}{l}{SUSI (nm)} & \multicolumn{1}{l}{TuMag}}
	\startdata
	Quiet Sun$^*$ & $01\_$QSUN &- & 2024-07-10 13:16 - 14:52  & 0.99 & 327-329 & \ion{Mg}{1}\,b$_2$ + \ion{Fe}{1} \\
	Emerging flux$^*$ &$02\_$EMEF$^{\dagger\diamond\oplus}$& 13738 & 2024-07-10 19:14 - 23:22& 0.97 &\ion{Ca}{2}\,K \& 407-410 & \ion{Mg}{1}\,b$_2$ + \ion{Fe}{1}  \\
	%& & & & & \\
	%
	Sunspot & $03\_$SPOT&13738 & 2024-07-11 02:22 -  02:34  & 0.97 & -& \ion{Mg}{1}\,b$_2$ + \ion{Fe}{1} \\
	Quiet Sun$^*$&$04\_$QSUN$^{\dagger\oplus}$ & - &2024-07-11 03:58 - 10:51 & 0.99 & \ion{Ca}{2}\,H\,K \& 407-410 & \ion{Mg}{1}\,b$_2$ + \ion{Fe}{1} \\
	Quiet Sun$^*$ &$05\_$QSUN$^{\dagger\diamond}$& - &2024-07-11 14:59 - 22:11 & 0.98 & FSS$^2$ & \ion{Fe}{1}\\
	%& & & & & \\
	%
	Sunspot$^*$ & $06\_$SPOT$^{\oplus}$&13743 & 2024-07-12 09:27 - 14:26& 0.88 &\ion{Ca}{2}\,K \& 407-410& \ion{Mg}{1}\,b$_2$ + \ion{Fe}{1}\\
	Plage  & $07\_$PLAG&13742 & 2024-07-12 15:46 -  15:50& 0.86 &\ion{Ca}{2}\,K & -\\
	Quiet Sun$^*$ &$08\_$QSUN& - &2024-07-12 19:16 -  19:46& 0.99 &406-408& \ion{Fe}{1} \\
	Sunspot$^*$ & $09\_$SPOT&13738 &2024-07-12 20:58 -  21:30 & 0.86 & 406-408&\ion{Fe}{1}\\
	%& & & & &\\
	%
	Flare$^*$ & $10\_$FLAR$^{\oplus}$&13738 &2024-07-13 00:59 -  03:21& 0.81 &  \ion{Ca}{2}\,H \& K & \ion{Mg}{1}\,b$_2$ + \ion{Fe}{1}\\
	Sunspot$^*$& $11\_$SPOT&13738 &2024-07-13 03:59 - 07:34& 0.80  &\ion{Ca}{2}\,H \& 406-410& \ion{Mg}{1}\,b$_2$ + \ion{Fe}{1}\\
	Flare$^*$ & $12\_$FLAR&13738 &2024-07-13 11:44  - 14:32 & 0.76& Part of FSS & \ion{Mg}{1}\,b$_2$ + \ion{Fe}{1} \\
	Flare$^*$ & $13\_$FLAR&13738 &2024-07-13 15:23  -  15:40& 0.76& \ion{Ca}{2}\,K & \ion{Mg}{1}\,b$_2$ + \ion{Fe}{1}\\
	Plage$^*$& $14\_$PLAG&13750 &2024-07-13 17:34  - 19:21& 0.29& \ion{Ca}{2}\,K &\ion{Mg}{1}\,b$_2$ + \ion{Fe}{1}\\
	Quiet Sun $^*$&$15\_$QSUN& - & 2024-07-13 19:42  - 21:08 & 1.00 & \ion{Ca}{2}\,K& - \\
	Limb &$16\_$LIMB& - &2024-07-13 21:52 - 21:55 & 0.15 & \ion{Ca}{2}\,K &-\\
	Limb$^*$&$17\_$LIMB& - &2024-07-13 22:05 - 00:10 (+1) & 0.25 & \ion{Ca}{2}\,K& \ion{Mg}{1}\,b$_2$ + \ion{Fe}{1}\\
	%& & & & &\\
	%
	Sunspot$^*$ & $18\_$SPOT&13738 &2024-07-14 00:32 - 00:47& 0.68& 406-409&\ion{Mg}{1}\,b$_2$ + \ion{Fe}{1}\\
	Plage$^*$ & $19\_$PLAG &13742 &2024-07-14 02:25 - 04:16 & 0.96& \ion{Ca}{2}\,K \& 407-410&\ion{Mg}{1}\,b$_2$ + \ion{Fe}{1} \\
	Limb$^*$ & $20\_$LIMB&- &2024-07-14 08:24 -  09:00 & 0.43 & Multiple regions &\ion{Mg}{1}\,b$_2$\\
	Plage & $21\_$PLAG& 13748 &2024-07-14 10:49 -  12:36 & 0.73 & 327-329 & \ion{Mg}{1}\,b$_2$ + \ion{Fe}{1}\\
	Sunspot$^*$&$22\_$SPOT$^{\dagger}$ & 13743 &2024-07-14 13:34 - 17:24 & 0.97 & Part of FSS &  \ion{Fe}{1} \\
	Quiet Sun$^*$ & $23\_$QSUN&- &2024-07-14 21:07 - 21:08 & 0.96 & 407-410 & -\\
	Sunspot$^*$ & $24\_$SPOT$^{\dagger\bullet\oplus}$&13744 &2024-07-14 21:26 - 00:36 (+1)& 0.95 & 407-410 \& 327-329 & \ion{Fe}{1}\\
	%Sunspot$^*$ & $25\_$SPOT$^{\dagger\bullet}$&13744 &2024-07-14 21:39 - 00:36 & 0.96 &  327-329 &  \ion{Fe}{1}\\
	%& & & & &\\
	%
	Limb$^*$&$25\_$LIMB& - &2024-07-15 02:41 -  03:53 & 0.37& 407-410 \& 358-360& \ion{Mg}{1}\,b$_2$ + \ion{Fe}{1}\\
	Sunspot&$26\_$SPOT& 13753 &2024-07-15 04:39 -  04:40 & 0.97& 407-410 \& 358-360& - \\
	Quiet Sun$^*$ & $27\_$QSUN&- &2024-07-15 05:04 - 07:53 & 0.97&  -& -\\
	Sunspot &$28\_$SPOT &13738 &2024-07-15 08:46 -  08:55& 0.36& 311-313& \ion{Mg}{1}\,b$_2$ + \ion{Fe}{1}\\
	Emerging flux&$29\_$EMEF$^{\dagger\diamond\bullet}$& - & 2024-07-15 10:08 -  14:04& 0.96 & 407-410& \ion{Mg}{1}\,b$_2$ + \ion{Fe}{1} \\
	%Quiet Sun&$31\_$QSUN& - & 2024-07-15 15:16 -  15:17& 0.94 & ?& - \\
	Plage$^*$ &$31\_$PLAG& - &2024-07-15 16:45 - 18:51& 0.97&  \ion{Ca}{2}\,K \& 407-410 & \ion{Mg}{1}\,b$_2$ + \ion{Fe}{1} \\
	Quiet Sun &$32\_$QSUN&- &2024-07-15 19:56 -  20:43& 0.99 & \ion{Ca}{2}\,K & \ion{Fe}{1}\\
	%Quiet Sun$^*$ &$34\_$QSUN&- &2024-07-15 20:04 -  20:43& 0.99 & \ion{Ca}{2}\,K & \ion{Fe}{1}\\
	AR filament & $33\_$ARFL$^\oplus$&13745 & 2024-07-15 22:05 -  22:42 & 0.96 & \ion{Ca}{2}\,K&\ion{Mg}{1}\,b$_2$ + \ion{Fe}{1} \\
	Quiet Sun &$34\_$QSUN& - & 2024-07-15 22:54 - 22:55& 0.97 & 311-313& -\\
	%& & & & &\\
	%
	Coronal hole$^*$ &$35\_$CHOL$^{\oplus}$& - & 2024-07-16 00:30 - 05:39& 0.87 & Multiple regions &\ion{Mg}{1}\,b$_2$ + \ion{Fe}{1}\\
	Flare & $36\_$FLAR& 13738 &2024-07-16 07:39 -  08:30& 0.29& 407-410 &\ion{Mg}{1}\,b$_2$ + \ion{Fe}{1}\\
	Limb$^*$ & $37\_$LIMB&- & 2024-07-16 11:09  - 13:57 & 0.29&  407-410 \& 386-388&\ion{Mg}{1}\,b$_2$ \\
	\enddata
	\tablenotetext{ }{\begin{itemize}
			\item[$^*$] Composed of multiple shorter duration observations between which at least one of the three instruments changed their observing mode or spectral region. 
			\item[$^1$]SCIP observed the \ion{Ca}{2} 854.2 nm and 849.8 nm, \ion{Fe}{1} lines around 850 nm, \ion{K}{1} D$_1$ \& D$_2$ and \ion{Fe}{2} 771.2 nm lines.
			\item[$^2$]SUSI Full Spectral Scan (FSS) extends from 309 nm to 417 nm covered by 40 different spectral windows.
			\item[$^\dagger$] Long timeseries from TuMag($>$120 minutes)
			\item[$^\diamond$] Long timeseries from SCIP($>$120 minutes)
			\item[$^\bullet$] Long timeseries from SUSI($>$120 minutes)
			\item[$^\oplus$] Contains SUSI and SCIP synchronous scans
		\end{itemize}
	}
\end{deluxetable*}

%\input{data_reduction}
%!TEX root = main.tex

\section{Data reduction}

\subsection{SCIP} 
The SCIP spectropolarimetric data acquired in the normal (standard) observation mode are compressed using a bit-compression scheme after onboard demodulation to reduce the data volume. Data reduction begins with bit-decompression, followed by the calibration procedures implemented in the SCIP data reduction pipeline. These procedures include bias subtraction, image skew correction, flat-field correction, polarisation calibration, and spatial alignment between the 770\,nm and 850\,nm wavelength channels. 

The image skew correction applies an image transformation that removes the spectral line curvature while ensuring uniform spatial and spectral sampling and fixed spectral-line positions. The transformation matrix is derived from flat-field measurements obtained at the time closest to each observation. Consequently, drift of the spectral positions during the slit-scanning is to be corrected in the released data, but has not yet been implemented in the current pipeline. The flat field for the 850\,nm wavelength channel is constructed from a combination of the flat-field data obtained during the flight, a spectral atlas, and the pre-flight test data. In contrast, we do not use a spectral atlas for the 770\,nm wavelength channel because numerous telluric lines appear in the atlas in this wavelength range. Instead, we perform polynomial fitting for the continuum spectra. The polarisation calibration includes correction of the rolling-shutter effect of the CMOS sensor, application of the inverse of the polarisation response matrix derived from pre-flight polarisation calibration \citep{kawabata2022}, merging of two orthogonal polarisation states and correction of residual crosstalk. The calibrated SCIP data achieve polarisation sensitivity of 0.04\%\ of the continuum intensity ($1\sigma$) in the 850 nm wavelength channel and 0.03\%\ ($1\sigma$) in the 770\,nm wavelength channel range with a 10 s integration in a quiet-Sun region near disc centre \citep{kubo2026}. 

The calibration procedures for spectroscopic data obtained in the rapid observation mode are the same as those of the standard mode, except that bit-decompression and polarisation calibration are not applied. Instead, an additional correction is required to remove the intensity modulation caused by the continuously rotating waveplate. The calibration of SJ imager data consists of bias subtraction and flat-field correction. The SJ image intensities are also modulated by the rotating waveplate, and this modulation is corrected in the pipeline.

\subsection{SUSI}
Here we provide a short overview of the data reduction strategy adopted for SUSI. A more detailed description of the final version, including a polarization sensitivity analysis, will be published in \cite{iglesiasetal2026}. The SUSI pipeline handles a large data volume (57 h/103 TB of science observations, 30 h/53 TB of in-flight calibration, 12 h/20 TB of pre-flight polarimetric calibration), so it relies heavily on post-flight parallel processing and GPU-based image restoration \cite[see][]{felleretal2025}. Across its 108-nm operating range, SUSI observed 47 different spectral windows that show intensity flux levels differing by an order of magnitude and strongly different levels of instrumental polarisation. Thermal changes during the flight introduced a slow image-registration drift ($\approx$2\,px/h), mitigated by tracking using the frequent in-flight calibrations taken at most within ~3 h of the corresponding science observation. The fine spatial sampling and optical layout made the system particularly sensitive to instrumental jitter, requiring precise dual-beam channel alignment and exploitation of the high frame rate with state-of-the-art image restoration, to restore image contrast and reduce jitter induced spurious polarimetric signals.

The main reduction steps applied to SUSI data are listed below.
(a) \textit{Reduction of SJ images}: This includes corrections for camera dark and bias, and the system flat field. The latter accounts for both static and time-varying components;
(b) \textit{Restoration of SJ images:} This step includes the co-registration of the two phase-diversity channels; and their image restoration, via the Multi-Object Multi-Frame Blind Deconvolution (MOMFBD) implementation by \cite{vannoortetal2005}, to  considerably reduce the effects of jitter and residual optical aberrations. The restoration results in one high-resolution restored SJ image every 3.07\,s; 
(c) \textit{Reduction of spectrograph (SP) intensity images:} Includes corrections for camera dark and bias; image distortion in the slit direction using the nearest grid target measurement; spectral line curvature (smile) correction and determination of the wavelength axis using the technique implemented by \cite{hoelkenetal2023} on the nearest flat-field measurement, while accounting for static and time-varying flat-field components \cite[see][for details]{felleretal2025};
(d) \textit{Restoration of SP images:} Each individual SP image is restored using the co-temporal point-spread functions derived from the previous restoration of SJ images. This very  computing intensive task is done using the technique and software recently developed by \cite{vanNoort17}. See also \cite{vannoortetal2025} for a recent demonstration;
(e) \textit{Stokes parameters derivation:} Includes the polarimetric demodulation of both channels \cite[see][for details]{iglesiasetal2025}, and the ad-hoc correction of residual polarimetric cross-talk \cite[following][]{jaegglietal2022} and other artifacts (e.g. polarised fringes). Note that the demodulation is done on restored SP images to obtain restored polarisation maps. The ad-hoc correction is accomplished by exploiting known polarisation properties of solar signals, such as the low correlation between Stokes~$I$ and the other Stokes parameters, the negligible polarisation levels of the continuum at disk center, and magnetically-insensitive spectral lines. 

\subsection{TuMag}

The reduction of TuMag data follows the standard steps of any solar spectropolarimeter: dark-current subtraction, flat-field correction, data demodulation, cross-talk removal, and wavefront reconstruction. However, the in-flight performance of the instrument depends critically on four aspects of its optical and thermal behavior:
\begin{itemize}
	\item the behavior of the Fabry–Pérot etalon, operated in a double-pass collimated configuration;
	\item the thermal evolution and band-pass stability of the pre-filters;
	\item the overall thermal stability of the instrument during the flight;
	\item the precise relative alignment of the two cameras forming the dual-beam system.
\end{itemize}

TuMag acquired an extensive set of calibration observations during the flight.
The mean and RMS values of the dark frames recorded by the cameras exhibited no significant variations during the flight, so a simple master-dark subtraction is sufficient and no additional correction is required.

The construction of the flat fields required a special treatment because of the double-pass collimated etalon configuration. Light coming to each pixel illuminates the etalon at a slightly different incidence angle, producing a field-dependent spectral blueshift and changes in the effective transmission. Furthermore, the pre-filters exhibited a wavelength shift with temperature. This shift was slow during the mission but still significant enough to be taken into account. For these reasons, flat fields cannot be applied directly. Instead, the reduction pipeline:
\begin{enumerate}
	\item measures a pixel-by-pixel blueshift map from dedicated flat-field calibration sequences;
	\item characterises the pre-filter transmission and its thermal drift;
	\item constructs a modified flat field by correcting the measured flats for blueshift and pixel-to-pixel gain inhomogeneities, and scaling them by the wavelength-dependent pre-filter transmission.
\end{enumerate}

This procedure avoids artificial gradients in intensity and spurious velocity patterns that arise when using the raw flat fields.

The alignment between the cameras is critical for accurate polarimetry. During the flight the cameras exhibited a small relative rotation up to $\sim0.1^\circ$, which is sufficient to generate artificial polarisation signals at the corners of the field of view. Using dedicated pinhole observations, we determined the relative shift and rotation between the cameras for each timeline. The measured offsets were small (on the order of a few pixels), but the rotation changed during the flight, likely due to thermal coupling with the radiators. The alignment is thus corrected on a timeline-by-timeline basis.

The instrument exhibited cross-talk from Stokes~$I$ into Stokes~$Q$ at a level of a few percent. This contamination is caused by the combined effect of jitter and non-ideal conditions during the ground calibrations, done at 1 bar while TuMag was defocused \citep{bailenetal2026}. The cross-talk was successfully removed using the method of \cite{jaegglietal2022}, which models the contamination as the action of a diattenuator on the Stokes vector. The parameters of the diattenuator are fitted independently for each dataset by minimizing the correlation between Stokes~$I$ and the rest of the Stokes parameters.

Each observing timeline was bracketed by PD acquisitions, which allowed us to monitor any possible changes in the optical wavefront during each timeline. No significant degradation was detected during the flight; nevertheless, the reconstruction uses the PD calibration closest in time to each science sequence. 
TuMag data display interference fringes. The probable source of one fringe pattern is the protective window of one of the cameras. Irrespective of their origin, the fringes are corrected. 

In summary, after reduction, TuMag data meet the stability and polarimetric accuracy required for scientific exploitation. At the time of writing, only TuMag data obtained in vector mode (measuring the four Stokes parameters) have been processed.  

The TuMag data reduction steps are to be described in detail by Orozco Suárez et al.\ (2026).

%\input{sample_data}
%!TEX root = main.tex
\section{Sample data}

Here we present sample datasets recorded by the three science instruments onboard \sunriseiii. 

\begin{figure*}
	\centering
	\includegraphics[width=1\textwidth]{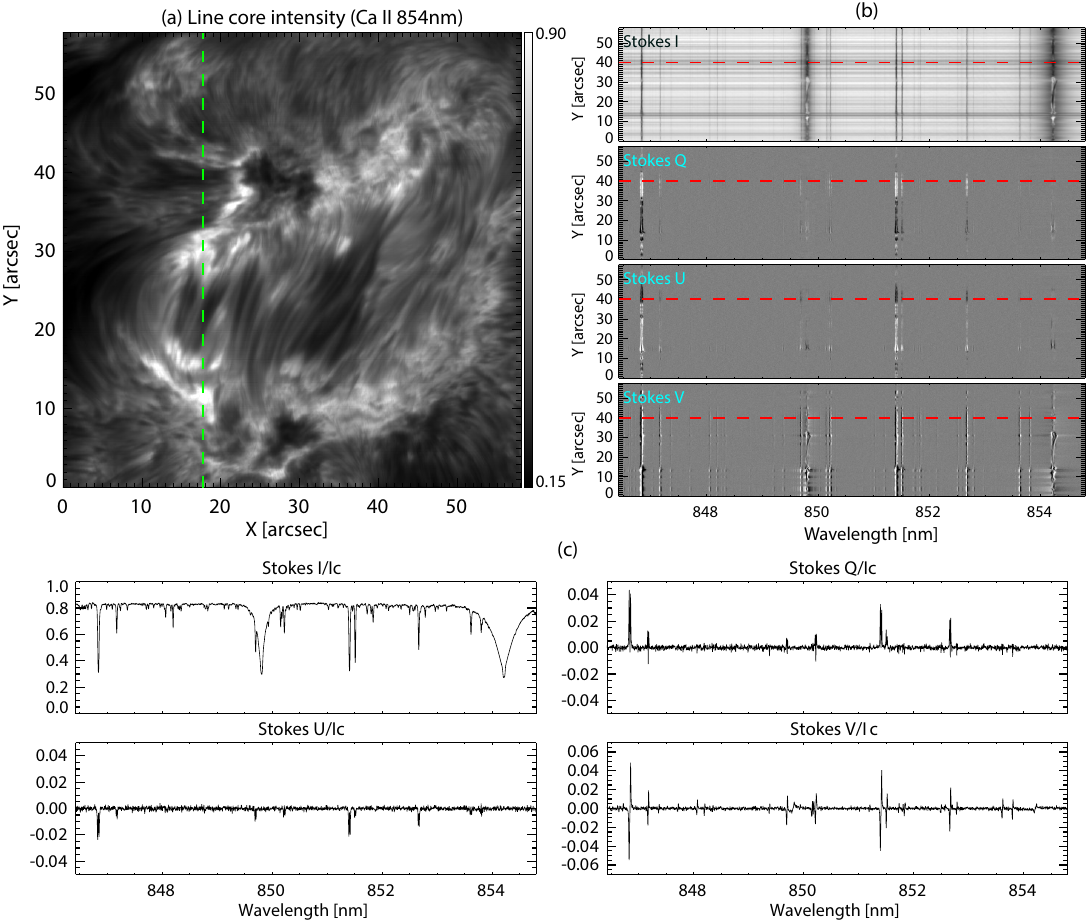}
	\caption{Example of SCIP data: (a) \ion{Ca}{2} 854 nm line-core intensity map of the emerging active region observed on 15 July 2024 (see Sunrise ID $30\_$EMEF in Table~\ref{tab:table}). (b) Stokes $I, Q, U,$ and $V$ data along the slit indicated by the green dashed line in panel (a). (c) Stokes profiles along the red dashed line in panel (b) normalized by the continuum intensity averaged over the quiet area in the image. The integration time of this dataset is 1 s at each slit position.}
	\label{fig:scip_data}
\end{figure*}

Figure~\ref{fig:scip_data} displays part of the data recorded by SCIP in a forming active region showing copious amounts of emerging flux on 2024 July 15. Panel (a) displays a full spatial scan as seen in the core intensity of the \ca~854~nm line. The cool loops forming an arch filament system associated with the flux emergence are prominently visible. Plotted in panels (b) are the Stokes parameters along the slit when it was located at the dashed green line in panel (a). Only the 850\,nm channel of SCIP is plotted. The two \ca~IR spectral lines are the most prominent in Stokes $I$, while in the polarised Stokes profiles the Fe~{\sc i}~846~nm line shows the strongest signals. Finally, in panels (c) the spectrum in a single pixel is shown, over the same wavelength range as in panel (b). It refers to the red dashed lines in panels (b).

\begin{figure*}
	\centering
	% norm to Ilocal
	\includegraphics[width=1\textwidth]{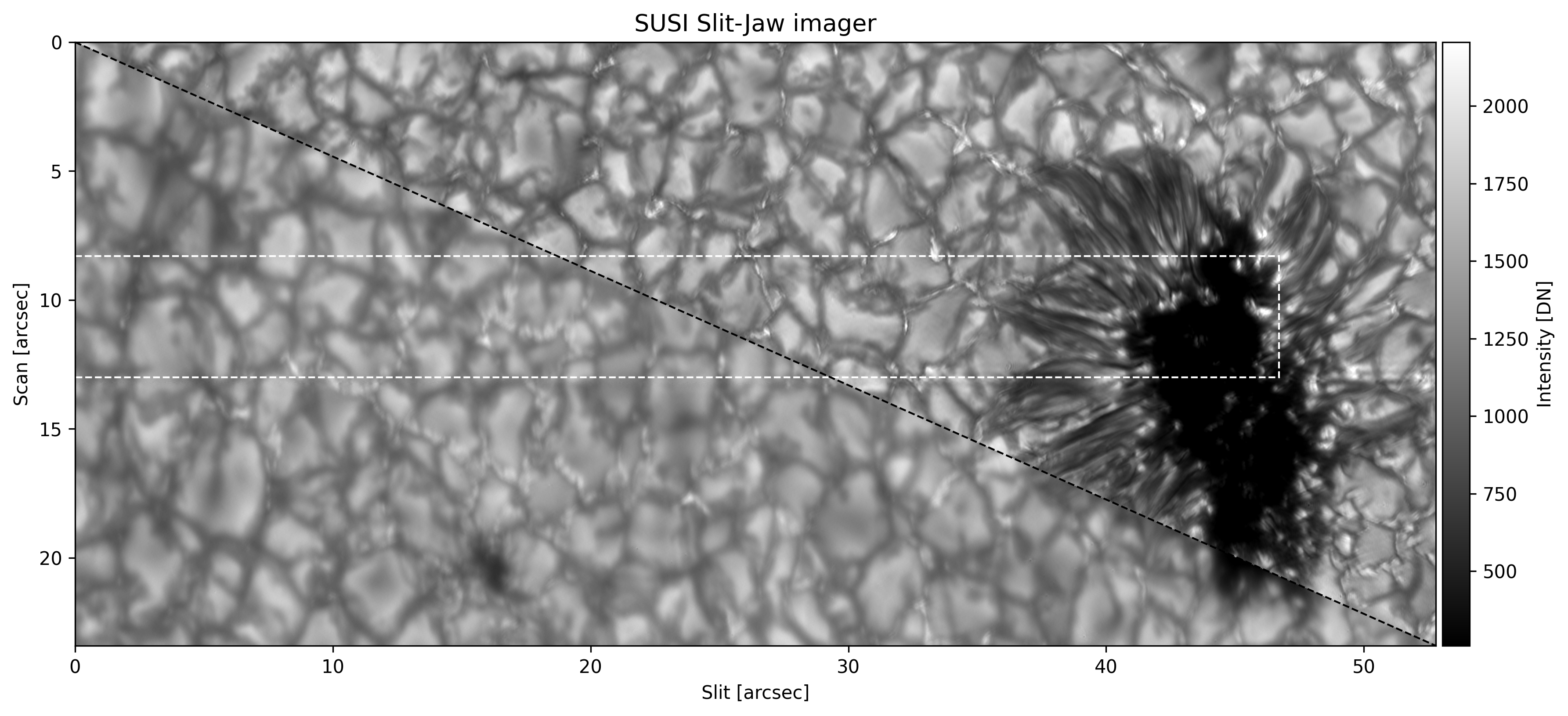}
	\includegraphics[width=1\textwidth]{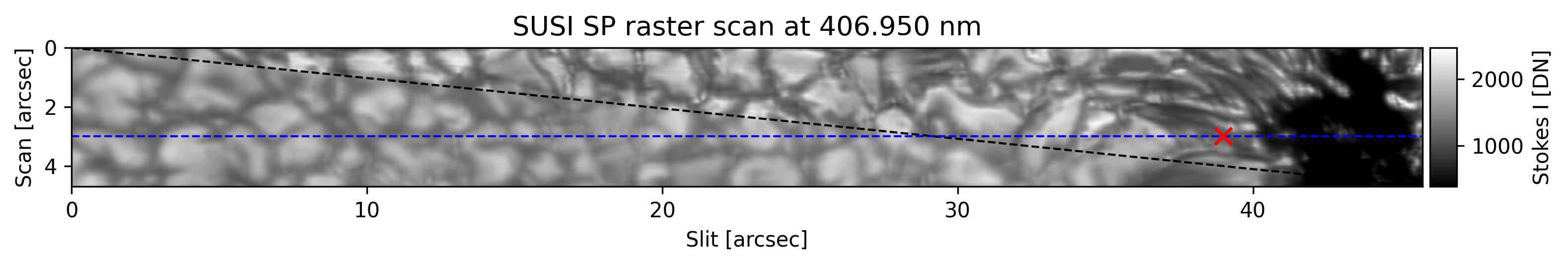}
	\includegraphics[width=1\textwidth]{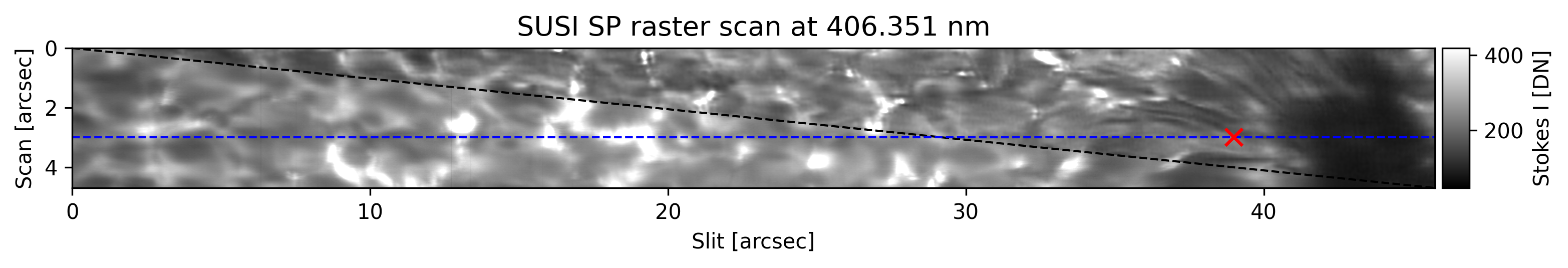}
	\caption{SUSI observation centered at 407.0~nm, acquired during a small 8\,min-long portion of the full spectral scan (FSS) on 2024-07-14 (see Sunrise ID 22\_SPOT in Table~\ref{tab:table}). Upper panel: a single slit-jaw image acquired at 17:16:26~UT, along with the co-temporal restored frame, separated by a black dashed line. The dashed white rectangle marks the approximate area scanned by the slit. Note that the slit, originally visible near the bottom of the white box, was removed using a Fourier filter. Middle panel: a single wavelength point at 406.950~nm (see the green lines in Figure~\ref{fig:susi_data2}) of the raster scan, recorded from 17:08:42 to 17:17:01 UT. Versions before and after image restoration are separated by the diagonal black dashed line. The dashed blue line and red cross mark the cut locations used for the slit spectra and single spectrum shown in Figure~\ref{fig:susi_data2}, respectively. {The pixel scales along the slit and scan directions are $0.0297"$ and $0.0307"$, respectively. Note that the axes scale in the top and middle panels are different.} Bottom panel: Same as middle panel but for wavelength 406.351\,nm }
	\label{fig:susi_data1}
\end{figure*}

\begin{figure*}
	\centering
	\includegraphics[width=1\textwidth]{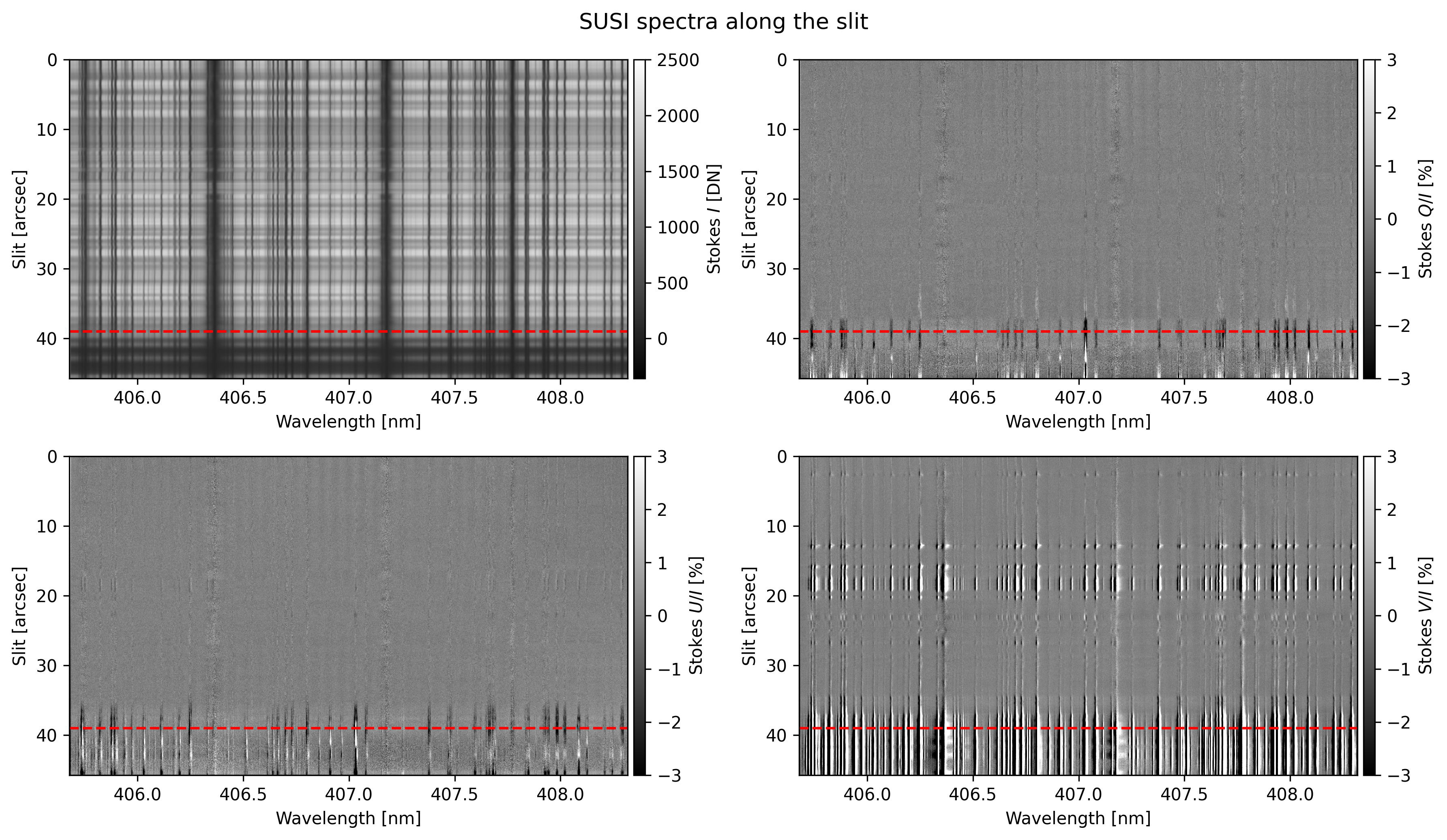}
	\includegraphics[width=1\textwidth]{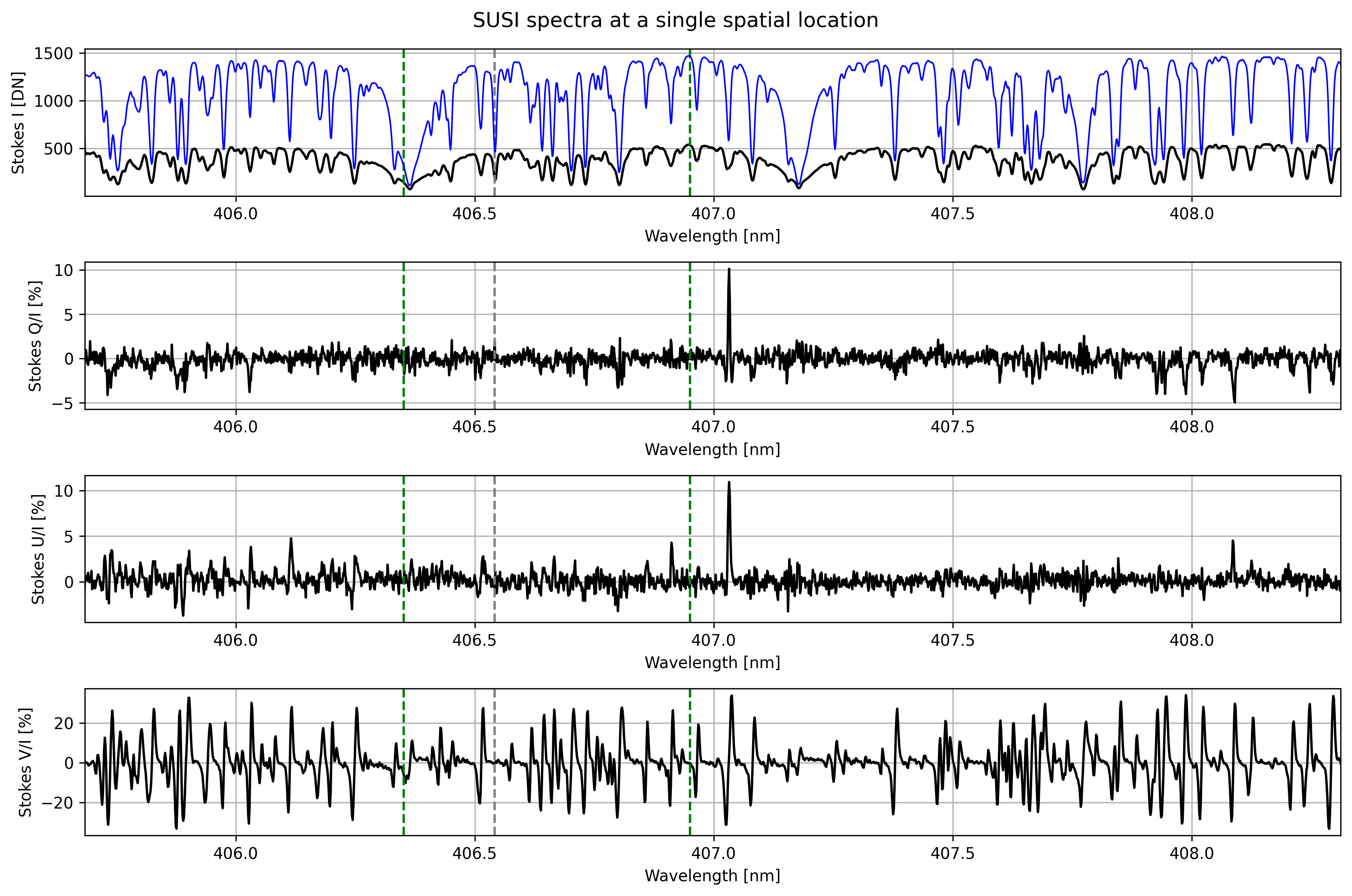} 
	\caption{Cuts through the non-restored SUSI observation presented in Figure~\ref{fig:susi_data1}. The top four images show the Stokes {spectrograms} along the slit (see blue line in Figure~\ref{fig:susi_data1}). The grey scale has been chosen to reveal the many weaker signals present in the data. The red dashed lines mark the spatial location (see also the red cross in Figure~\ref{fig:susi_data1}) corresponding to the bottom four spectra shown in black. The blue spectrum shown in Stokes I corresponds to the average along the full slit. The green dashed lines mark the wavelength positions of the raster scans shown in Figure~\ref{fig:susi_data1}. The gray dashed line marks the Fe I absorption line at 406.538\,nm, which has zero effective Lande g-factor. {The spectral line with the strongest $Q/I$ signal is \ion{Mn}{1} 407.028 nm}. Residual instrumental effects may still be present at this stage of the data reduction.}
	\label{fig:susi_data2}
\end{figure*}

The top panel of Figure~\ref{fig:susi_data1} shows an image taken by the slit-jaw imager on SUSI. The bottom left half of the image displays the reduced but not phase-diversity restored data, while the upper right half depicts the additionally restored data, highlighting the improvement in contrast. The two lower panels result from spectrograph scans. They correspond to a single wavelength each (given above the panels) and cover the spatial extent of the dashed white rectangle in the top panel. They too highlight the difference between the unrestored and restored data. The wavelength 406.95~nm corresponds to a near-continuum intensity, while 406.351~nm to near the core of a strong line. These wavelengths are indicated by the green dashed lines in the bottom 4 panels of Fig.~\ref{fig:susi_data2}.

Figure~\ref{fig:susi_data2} displays Stokes $I,Q,U,V$ spectra. In the top four panels these are plotted along the slit at one particular spatial position (blue dashed line in the lower two panels of Fig.~\ref{fig:susi_data1}). In the bottom four panels, in black the Stokes spectrum at a single pixel is shown (corresponding to the dashed red line in the upper four panels of this figure and to the red cross in Fig.~\ref{fig:susi_data1}). The blue Stokes $I$ spectrum is the average over the whole slit. The Fe~{\sc i}~406.538~nm (dashed grey line in the lower four panels) has a Land\'e factor of zero and as expected shows no signal above the noise in the polarised Stokes spectra. The opposite is the case for Mn~{\sc i}~407.028~nm, which was identified by \cite{1973SoPh...28....9H} as a line with a particularly large Land\'e factor of 3.33 and correspondingly displays the strongest Stokes $Q,U$ and $V$ signals in this wavelength range.  

\begin{figure*}
	\centering
	\includegraphics[width=1\textwidth]{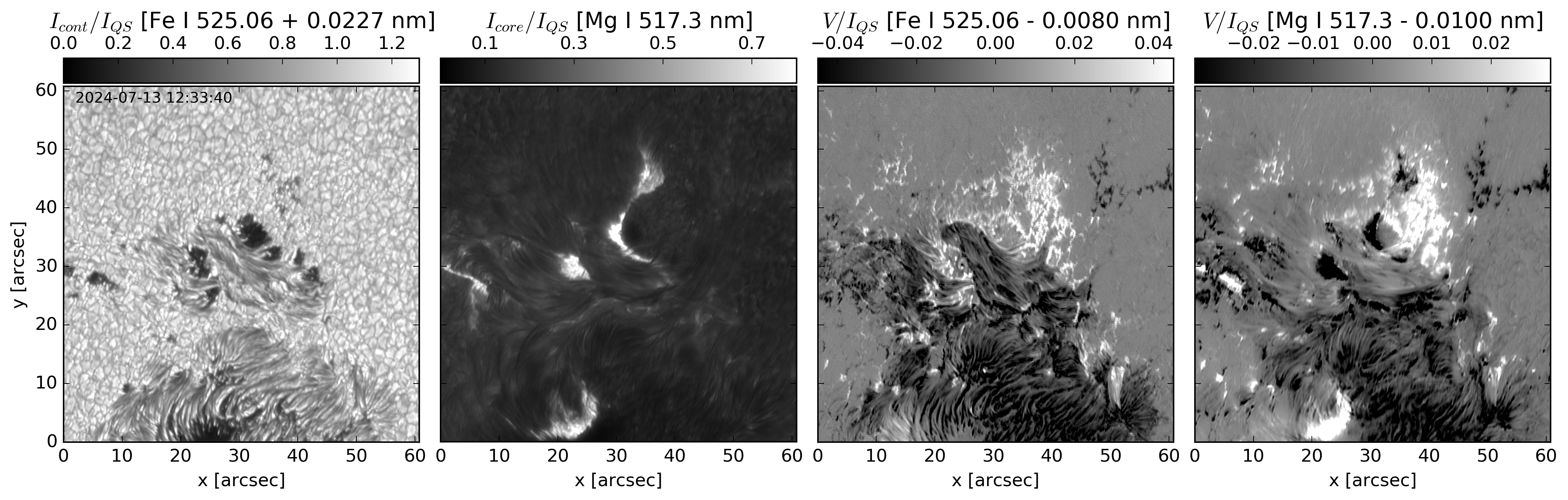}
	\caption{Example TuMag observations obtained on 13 July 2024 during an M5.3-class flare in NOAA AR 13738, showing a frame close to the flare peak. The panels display, from left to right: (1) the continuum intensity at $+22.7$\,pm ($+227$\,m\AA) from the \ion{Fe}{1} 525.06\,nm line, (2) the \ion{Mg}{1} 517.26\,nm line-core intensity, (3) Stokes $V$ in the \ion{Fe}{1} 525.06\,nm line wing at $-8$\,pm ($-80$\,m\AA), and (4) Stokes $V$ in the \ion{Mg}{1} 517.26\,nm line wing at $-10$\,pm ($-100$\,m\AA). All panels are normalized to the average quiet-Sun continuum intensity, $I_{QS}$. The data have been reconstructed using wavefront/phase-diversity techniques. Residual instrumental effects may still be present at this stage of the data reduction. %\hns{Ishikawa-san suggests rotating the maps by 180 deg so that top is north. Should we ask Azaymi to rotate the figure? } \todo{Sami: Yes, please do. Thank you!}
	}
	\label{fig:tumag_data}
\end{figure*}

An example TuMag dataset is exhibited in Fig. \ref{fig:tumag_data}. It shows an M5.3-class flare taking place in and around a sunspot within NOAA AR 13738. It shows both the continuum intensity and the intensity in the line core of the Mg~{\sc i}~517.26~nm line, which is formed in the low chromosphere (the two left frames) as well as the Stokes $V$ signals in the Fe~{\sc i}~525.06~nm and the Mg~{\sc i}~517.26~nm line (two panels on the right). Note the fine-structure in the Stokes $V$ signals and in the flare ribbons. 
These images are part of a 2\,h 48\,min long time series covering this flare (the observation with Sunrise\_ID 12\_FLAR in Table~\ref{tab:table}).

%\input{science_results}
%!TEX root = main.tex

\section{First science results }

A sample of first science results obtained with data from the 2024 flight of \sunriseiii\ are very briefly described below. These are all described in much more detail in the other papers of this focus issue. 

Elongated structures with weaker or even opposite polarity line-of-sight chromospheric magnetic ﬁeld are found in the canopy fields extending from  network elements. These have no obvious photospheric counterpart directly below them and are associated with up- and downflows at their sides \citep{kubo2026}. 

An MHD simulation with the chromospheric extension of the MURaM code \citep{przybylskietal2022} reveals that these elongated structures  of  opposite polarity field are signatures of twisted magnetic flux ropes that are associated with vortical flows \citep{Ondratscheketal2026}.

Line-of-sight magnetic fields in spicules measured with the \ca~8542~\AA\ line are consistent with lower resolution vector field measurements in the He~{\sc i}~10830~\AA\ triplet, but show significantly weaker fields than found from ground-based measurements in the same line \citep{naitoetal2026}.

Ca~{\sc ii}~K profiles recorded by \sunriseiii/SUSI have been used to find relationships to convert historical solar images recorded in this spectral line in different pass-bands to a common scale. The relationships were applied to sample images and shown to greatly increase the consistency between datasets from different observatories \citep{yadavetal2026}.

The first detection of linear polarisation within an active-region  filament in the \ca\,854.2\,nm line is reported by \citet{matsumoto2026}. The detected signals correspond to a transverse  field of 300-500~G along different stretches of the filament. 

Two Ellerman bombs were studied with SCIP data in an emerging flux region. Evidence of bi-directional reconnection outflows originating from the upper photosphere and lower chromosphere was found \citep{kawabataetal2026}.

Joint observations of an M-class flare with the Domeless Solar Telescope at Hida observatory revealed chromospheric flare kernels to show a fibrilar substructure with the asymmetries and strengths of chromospheric line profiles and their evolution being different from one fibril to another \citep{Asai2026}. 

Observations by the SUSI instrument confirmed the prediction by \citet{harnesetal2025} that the Stokes $V$ profile of the Fe\,{\sc i}~407.17\,nm line can show two lobes with the same sign in each wing. It was also demonstrated that this unique feature allows the measurement of the LOS magnetic field at two different heights using the weak-field-approximation \citep{harnesetal2026}.

The propagation of roughly 5-min period acoustic waves from the solar surface to the lower chromosphere was traced by measuring the line-core positions of 19 spectral lines in the 396.8\,nm region. The phases of the oscillations follow the formation heights of the spectral lines, and reveal up- and downward propagating waves with phase speeds of $>$10\,km\,s$^{-1}$ \cite[]{laggetal2026}.

A time series analysis of 44 spectral lines in the range 327-329~nm in a sunspot umbra reveals that each line shows its own unique signature of oscillation frequencies in the range 2-9\,mHz. The frequency pattern shows no clear dependence on the height of formation of the lines 
%as computed in a snapshot of a radiation-MHD sunspot simulation 
\citep{jafarzadehetal2026a}.

A further analysis of a subset of these lines across different solar features (sunspot umbra, penumbra, pore, plage, quieter region) uncovers clear differences in power in the 2--4~mHz, 4--6~mHz and 6--8~mHz, with stronger field regions showing depressed power in the low frequencies and enhanced power in the higher frequencies. \citep{jafarzadehetal2026b}.
%\todo{(Jafarzadeh et al. 2026b)}.

The bright chromospheric ribbons of an M-class flare are found to be spatially interrupted by chromospheric loops overlying them. The Mg~{\sc i} b spectral line indicates the absorption of radiation from the underlying bright ribbon in these dark loops. This observation reveals a new source of fine structure in flare ribbons \citep{chittaetal2026}.  

A flare ribbon of a C-class flare is found to show at most weak evidence of chromospheric evaporation, but instead indicates the presence of small-scale reconnection in low chromospheric layers. This indicates that in this particular case, the flaring mechanism deviates from the standard flare model \citep{QuinteroNodaetal2026}.%\sks{(Quintero Noda et al. 2026}.

Longer period (typically 5~min) waves are seen to propagate within small-scale magnetic elements in the quiet Sun internetwork from the photosphere to the chromosphere, where they produce clear signatures of shocks, while only higher frequency and lower amplitude waves reach the chromosphere outside these features \citep{obaetal2026}.%\todo{Oba et al. 2026}

The different polarity of Stokes V profiles of photospheric and chromospheric spectral lines provides evidence for magnetic reconnection occuring within a bald patch in the solar photosphere. This reconnection is proposed to trigger an M5.3 flare \citep{ishikawaetal2026}.

More than a hundred transitions, spanning elements from Hydrogen (H$\delta$) to rare-earth elements as well as molecular transitions were identified in off-limb spectra obtained by SUSI in its 409~nm window. Lines that are either extremely weak or completely blended on disc, become prominent off-limb \citep{CastellanosDuran2026a}. 

High-resolution multi-line observations of an emerging flux region obtained with \sunriseiii\ reveal strong plasma acceleration at the footpoints of an arch filament system. Combined measurements from TuMag and SCIP suggest a close relation between plasma dynamics and changes in magnetic connectivity \citep{gonzalezemanriquetal2026}.

 The thermal and magnetic properties of coronal rain resulting from a coronal condensation near a pore are constrained by SUSI. Whereas the temperature was stable to better than a factor of two, the magnetic field strength associated with the coronal rain varied by an order of magnitude \citep{kriginskyetal2026}.
	
	Off-limb observations of the \ca~K line show a broad line core in emission that remains visible up to 6\,Mm above the limb. Computations could reproduce the main properties of these off-limb observations only if stacked 3D MHD simulations are used and the curvature of the solar surface is taken into account \citep{milicetal2026}.

Note that the above list is incomplete as it only covers the papers submitted to this focus issue relatively early.

%\input{conclusions}
%!TEX root = main.tex

\section{Data access and conclusions}
The hardware phase of the \sunriseiii\ project culminated in a successful flight in 2024 which returned an order of magnitude more data than the earlier Sunrise~{\sc i} and {\sc ii} flights combined. Together with the more complex and capable instrumentation flying on \sunriseiii, this has produced an extremely rich set of data. The number of publications and the variety of topics covered in this Focus Issue of The Astrophysical Journal Letters is witness to this richness. 

At the time of writing, however, only a fraction of the gathered data have been fully reduced and an even smaller fraction have been analyzed. Consequently, there is far more science to be done with these data than that presented in this Focus Issue. All reduced data can be downloaded from the \href{https://sr3data.mps.mpg.de}{{Sunrise Mission Archive}}\footnote{{Sunrise Mission Archive}: \href{https://sr3data.mps.mpg.de}{https://sr3data.mps.mpg.de}}.

There was also an impressive number of coordinated co-observations with \sunriseiii\ done by almost all medium- and high-resolution ground-based solar telescopes and by all high-resolution solar space 
missions.\footnote{The exception is Solar Orbiter, which was on the far side of the Sun at the time of the \sunriseiii\ science flight.} See \citet{korpi-laggetal2025} for a list of participating observatories. These data, which complement the \sunriseiii\ data in different ways and have so far been analyzed to a very small extent only, represent an additional treasure trove for the future.

\begin{acknowledgements}
    Sunrise III is supported by funding from the Max-Planck-Förderstiftung (Max Planck Foundation), NASA under Grants \#80NSSC18K0934 and \#80NSSC24M0024 (“Heliophysics Low Cost Access to Space” program), and the ISAS/JAXA Small Mission-of-Opportunity program and JSPS KAKENHI Grant Numbers JP18H05234 and JP23K25916. This research has received financial support from the European Union’s Horizon 2020 research and innovation program under grant agreement No. 824135 (SOLARNET) and No. 101097844 (WINSUN) from the European Research Council (ERC). It has also been funded by the Deutsches Zentrum für Luft- und Raumfahrt e.V. (DLR, grant no. 50 OO 1608). The Spanish contributions have been funded by the Spanish MCIN/AEI under projects RTI2018-096886-B-C5, and PID2021-125325OB-C5, and from “Center of Excellence Severo Ochoa” awards to IAA-CSIC (SEV-2017-0709, CEX2021-001131-S), all co-funded by European REDEF funds, “A way of making Europe".  CK acknowledges grant RYC2022-037660-I funded by MCIN/AEI/10.13039/501100011033 and by ``ESF Investing in your future'' and grant PID2024-156066OB-C55, funded by MCIN/AEI/ 10.13039/501100011033 and by ``ERDF A way of making Europe’’.
\end{acknowledgements}

\bibliographystyle{aasjournalv7}
%\bibliography{main} 

%\input{appendix}

\end{document}